\documentclass[conference]{IEEEtran}
\IEEEoverridecommandlockouts
\usepackage{cite}
\usepackage{amsmath,amssymb,amsfonts}
\usepackage{algorithmic}
\usepackage{graphicx}
\usepackage{textcomp}
\usepackage{xcolor}
\usepackage[hyphens]{url}
\usepackage{fancyhdr}
\usepackage{hyperref}
\usepackage{braket}
\usepackage{subscript}
\usepackage{algorithm}
\usepackage{multirow}
\usepackage{makecell}
\usepackage[flushleft]{threeparttable}
\usepackage[caption = false]{subfig}

\usepackage{amsthm}

\newtheorem{lemma}{Lemma}
\newtheorem{prop}{Proposition}

\makeatletter
\newcommand{\newlineauthors}{%
  \end{@IEEEauthorhalign}\hfill\mbox{}\par
  \mbox{}\hfill\begin{@IEEEauthorhalign}
}
\makeatother

\def\BibTeX{{\rm B\kern-.05em{\sc i\kern-.025em b}\kern-.08em
    T\kern-.1667em\lower.7ex\hbox{E}\kern-.125emX}}

\begin{document}

\title{QuCLEAR: Clifford Extraction and Absorption for Quantum Circuit Optimization}

\author{
\IEEEauthorblockN{Ji Liu}
\IEEEauthorblockA{\textit{Argonne National Laboratory}\\
Lemont, USA \\
ji.liu@anl.gov}
\and
\IEEEauthorblockN{Alvin Gonzales}
\IEEEauthorblockA{\textit{Argonne National Laboratory}\\
Lemont, USA \\
agonzales@anl.gov}
\and
\IEEEauthorblockN{Benchen Huang}
\IEEEauthorblockA{\textit{University of Chicago}\\
Chicago, USA \\
benchenh@uchicago.edu}
\newlineauthors
\IEEEauthorblockN{Zain Hamid Saleem}
\IEEEauthorblockA{\textit{Argonne National Laboratory}\\
Lemont, USA \\
zsaleem@anl.gov}
\and
\IEEEauthorblockN{Paul Hovland}
\IEEEauthorblockA{\textit{Argonne National Laboratory}\\
Lemont, USA \\
hovland@mcs.anl.gov}
}

\maketitle
\thispagestyle{plain}
\pagestyle{plain}

\begin{abstract}
Quantum computing carries significant potential for addressing practical problems. However, currently available quantum devices suffer from noisy quantum gates, which degrade the fidelity of executed quantum circuits. Therefore, quantum circuit optimization is crucial for obtaining useful results. In this paper, we present QuCLEAR, a compilation framework designed to optimize quantum circuits. QuCLEAR significantly reduces both the two-qubit gate count and the circuit depth through two novel optimization steps. First, we introduce the concept of Clifford Extraction, which extracts Clifford subcircuits to the end of the circuit while optimizing the gates. Second, since Clifford circuits are classically simulatable, we propose Clifford Absorption, which efficiently  processes the extracted Clifford subcircuits classically. We demonstrate our framework on quantum simulation circuits, which have wide-ranging applications in quantum chemistry simulation, many-body physics, and combinatorial optimization problems. Near-term algorithms such as VQE and QAOA also fall within this category. Experimental results across various benchmarks show that QuCLEAR achieves up to a $77.7\%$ reduction in CNOT gate count and up to an $84.1\%$ reduction in entangling depth compared with state-of-the-art methods. 

\end{abstract}

\section{Introduction}
Quantum circuit optimization is important in both the Noisy-Intermediate-Scale-Quantum (NISQ) era and the fault-tolerant quantum computing era. Since quantum gates are noisy and the number of gates directly impacts circuit execution time, optimizing these circuits is essential. Given that operations on quantum devices have higher error rates than those on classical devices, we must ask: ``Do we need to run the entire circuit on quantum devices?" Hybrid algorithms such as VQE~\cite{peruzzo2014VQE} and QAOA~\cite{farhi2014qaoa} demonstrate that the evaluation of unitaries can be combined with classical optimizers on classical devices. In this paper, we show that quantum circuit size can be further reduced with circuit optimization and classical post-processing. We demonstrate the effectiveness of this approach on quantum simulation circuits.

Quantum simulation (also known as Hamiltonian simulation) is an important class of quantum algorithms. Given a Hamiltonian $H$ and evolution time $t$, the circuit implements the operator $U = e^{-i\hat{H}t}$. The operator is typically decomposed into the sum of local Hamiltonians~\cite{lloyd1996universal} with Trotterization~\cite{trotter1959trotter}, and the unitary is expressed in terms of a sequence of Pauli strings: $U =e^{-iP_1t_1}e^{-iP_2t_2}...e^{-iP_kt_k}$. In fact, any quantum circuit can be described through quantum simulation, since any unitary operation can be expressed in this form. Quantum simulation has numerous applications, including quantum chemistry~\cite{poulin2015trotter}, many-body physics~\cite{raeisi2012manybodyphysics, jiang2018manybodyphysics}, and solution of combinatorial optimization problems~\cite{guerreschi2019qaoa_maxcut}. For near-term algorithms such as VQE for chemistry simulation~\cite{grimsley2019trotterizedUCCSD} and QAOA for combinatorial optimization~\cite{khairy2020QAOA_combinatorial}, their ansatzes are generated by matrix exponentials, since these better describe the systems to be simulated or the problems encoded in the Hamiltonians. For instance, the recent demonstration of quantum utility~\cite{kim2023quantum_utility} is also based on this concept.

While quantum simulation is an important class of algorithms, the circuit consists of a relatively large number of basis gates. For example, the UCCSD ansatz~\cite{peruzzo2014VQE} for quantum chemistry simulation requires $O(n^4)$ Pauli strings, and the circuit is typically too large for running meaningful system sizes on current quantum devices. Quantum simulation circuits can be optimized by reordering the Pauli strings to maximize gate cancellation~\cite{li2022paulihedral, jin2023tetris}, simultaneously diagonalizing the building blocks to reduce gate count~\cite{cowtan2019pytket,cowtan2020UCCSD_diagonalization, van2020diagonalization}, or directly synthesizing a Pauli network~\cite{de2024rustiq}. The limitation of many existing works, however, is that they rely on local rewriting or synthesizing techniques without leveraging the properties of the quantum simulation circuits. Additionally, many of these optimizations preserve the unitary matrix of the circuit. In this work, we demonstrate that not all gate operations need to be executed on quantum devices. Specifically, operations such as Clifford gates can be efficiently processed classically. 

In this paper, we propose a framework that converts quantum simulation programs to hybrid programs, that is, converts a quantum circuit into a smaller quantum circuit with classical post-processing. This framework, named QuCLEAR (Quantum CLifford Extraction and AbsoRption), significantly reduces the number of two-qubit gates and circuit depth through two novel optimization techniques: Clifford Extraction and Clifford Absorption. Leveraging the properties of quantum simulation circuits, we propose Clifford Extraction which extracts the Clifford subcircuits to the end of the circuit while optimizing the circuit. The extracted Clifford subcircuits at the end can be efficiently simulated classically. We propose Clifford Absorption which processes the Clifford subcircuits classically by updating the measurement observables or efficiently post-processing the measurement probabilities. 
QuCLEAR follows a modular design that ensures its portability across different software and hardware platforms. Our experimental comparisons using comprehensive benchmarks demonstrate that QuCLEAR outperforms the state-of-the-art approaches in terms of both gate count and circuit depth.

Our contributions are as follows.
\begin{itemize}
\item We introduce QuCLEAR, a framework that optimizes quantum circuits. Experimental results across various benchmarks show that QuCLEAR achieves up to a $77.7\%$ reduction in CNOT gate count and up to an $84.1\%$ reduction in entangling depth compared to state-of-the-art methods.
\item We propose the technique Clifford Extraction which optimizes quantum circuits. Additionally, we introduce an efficient heuristic algorithm for synthesizing CNOT trees to be extracted in quantum simulation circuits.
\item We propose the technique Clifford Absorption which processes the Clifford gates at the end of the circuit classically by updating the measurement observables or post-processing the measurement probabilities.
\item We demonstrate that QuCLEAR is effective in optimizing logical circuits and may outperform hardware-aware algorithms when mapping to devices with limited connectivity (see Fig.~\ref{fig:sparse_backend}).
\end{itemize}


This paper is organized as follows. In Section~\ref{sec:background}, we introduce the background and the state-of-the-art works. In Section~\ref{sec:motivation}, we discuss the motivation and key observations. Section~\ref{sec:framework} introduces the high-level design of the QuCLEAR framework. In Section~\ref{sec:CE}, we present the algorithm for the Clifford extraction technique. In Section~\ref{sec:CA}, we present the proof and the algorithm for the Clifford absorption technique. In Section~\ref{sec:methodology}, we provide our experimental methodology. In Section~\ref{sec:evaluation}, we discuss our compilation results and benchmark comparisons. Section~\ref{sec:conclusion} provides a brief summary.

\section{Background}
\label{sec:background}
\begin{figure}[htbp]
\centerline{\includegraphics[width=\linewidth]{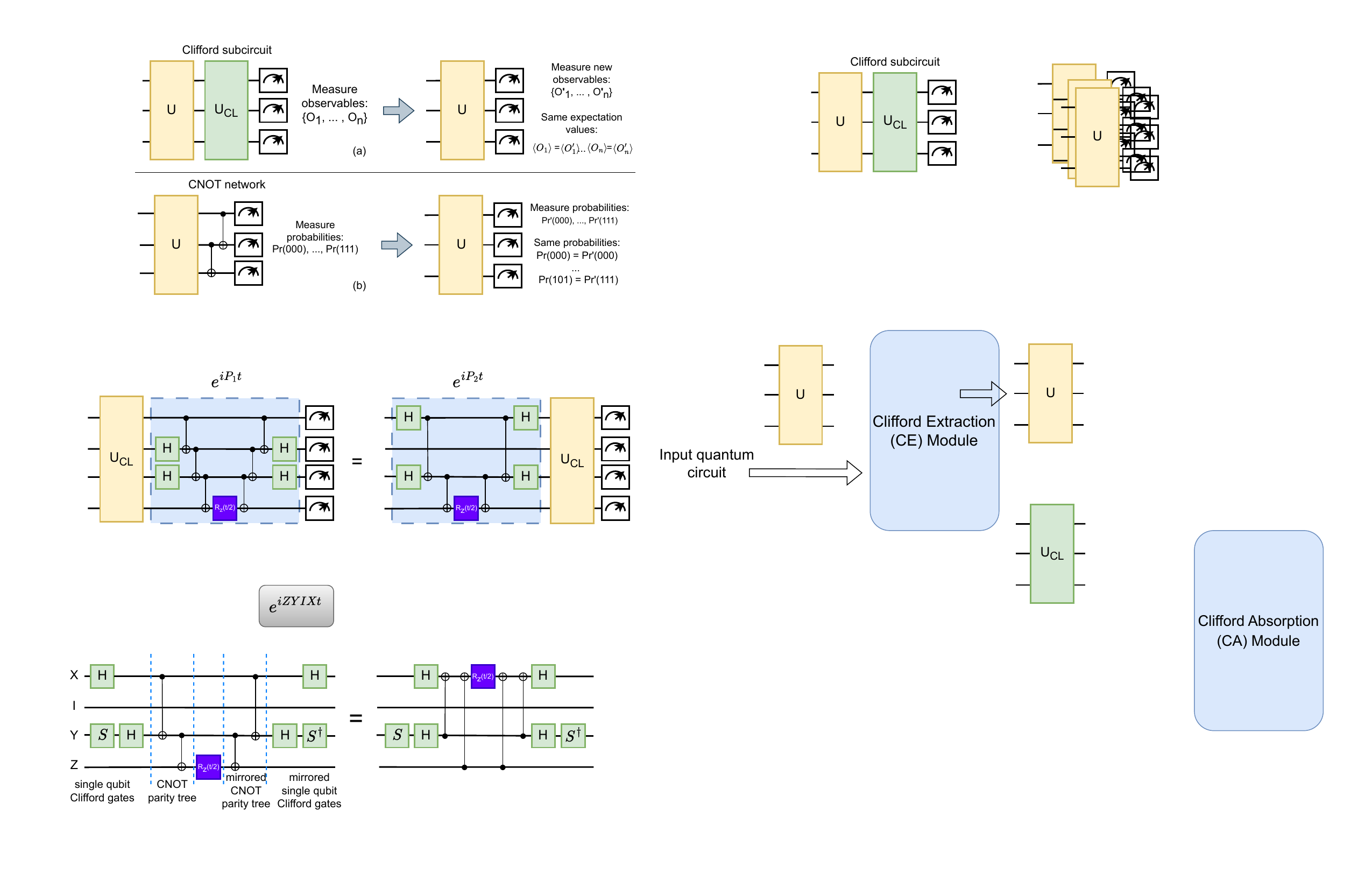}}
\caption{Equivalent quantum simulation circuits for $e^{iZYIXt}$}
\label{fig:HS_example} 
\end{figure}
\subsection{Quantum Simulation}
Quantum computation has the potential to benefit solving practical physics and chemistry problems~\cite{cao2019quantum}. Two major applications that have been studied are (i) solving the eigenvalue and eigenvector of (static) many-body Hamiltonian $\hat{H}$ and (ii) performing Hamiltonian simulation $e^{-i\hat{H}t}$. The former can be achieved using variational quantum algorithms by optimizing a parameterized quantum circuit, commonly referred to as an ansatz~\cite{peruzzo2014VQE}. One popular type of ansatz is the chemically -inspired ansatz~\cite{romero2018strategies}, which involves the exponentiation of qubit-encoded fermionic excitations, for example, $\Pi_{\text{exc}} e^{-i\theta P_{\text{exc}}}$, where $P_{\text{exc}}$ is the qubit operator (written in Pauli products) and $\theta$ is the variational parameter. The latter requires implementing $e^{-i\hat{H}t}$ on quantum circuits, which can be approximated through Trotterization $e^{-i\hat{H}t} \approx \left[\Pi_j e^{-i h_j P_j \Delta t}\right]^{t/\Delta t}$, where we have encoded the Hamiltonian into qubits as $\hat{H} = \sum_j h_j P_j$ and $\Delta t$ is the timestep. Interestingly, both solutions adopt quantum circuits that share a similar building block $e^{-i\theta P}$, that is depicted in Figure~\ref{fig:HS_example}. Moreover, a similar structure can be extended to QAOA optimization problems. Therefore, optimizing quantum circuits with this building block would benefit a broad range of applications and also improve the practicality of quantum computation on not only near-term but also fault-tolerant hardware. 

The buliding block circuit in Fig.~\ref{fig:HS_example} consists of several components. First, a layer of single-qubit Clifford gates handles the basis change: Hadamard gates or S gates followed by Hadamard gates are used to rotate the X and Y basis to the Z basis, respectively. Next, a layer of CNOT gates forms a CNOT parity tree, with the root qubit encoding the parity of all the qubits with non-identity Pauli operator~\cite{mukhopadhyay2023synthesizing_parity}. A single-qubit Rz gate is applied to the root qubit to introduce a phase shift. The circuit then mirrors the CNOT parity tree and single-qubit gates to perform the uncomputation. As shown in Fig.~\ref{fig:HS_example}, multiple valid circuit designs exist, as long as the Rz gate is applied to the root qubit that encodes the parity of all the qubits with non-identity Pauli operator. The total number of CNOT gates in a parity tree equals the number of non-identity Pauli operators minus one, since the qubit for each non indentity term in the Pauli tensor product (except for the root) serves as the control for one CNOT gate. 

\begin{figure*}[htbp]
\centerline{\includegraphics[width=\linewidth]{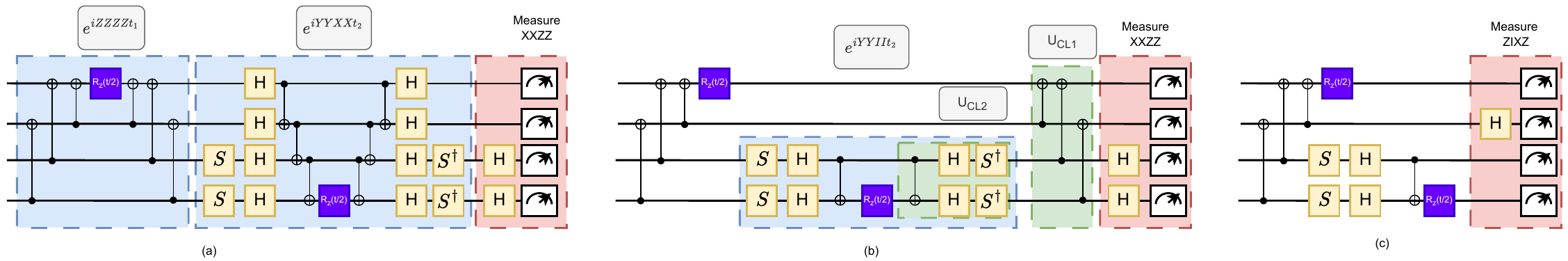}}
\caption{Optimizing quantum simulation circuit $e^{iZZZZt_1}e^{iYYXXt_2}$. (a) Original quantum simulation circuit, which can not be optimized with gate cancellation. The observable being measured is XXZZ. (b) Circuit after extracting the Clifford subcircuit in $e^{iZZZZt_1}$. The second Pauli rotation circuit is optimized to $e^{iYYIIt_2}$. (c) Circuit after absorbing the Clifford subcircuit in the observable measurement. The new observable is ZIXZ.}
\label{fig:opt_example} 
\end{figure*}
\subsection{Clifford Circuits}
Clifford circuits are a specific class of quantum circuits composed of operations from the Clifford group, which typically includes the Hadamard gate $H$, phase gate $S$, and CNOT gate~\cite{nielsen2010quantum}. 
A key feature of Clifford circuits is that they can be efficiently simulated on classical computers. The Gottesman-Knill theorem ~\cite{gottesman1998heisenberg} states that any quantum circuit consisting solely of Clifford gates can be simulated in polynomial time on a classical computer. Although Clifford unitary matrices are $2^n \times 2^n$ in size, they can be compactly represented by using the stabilizer tableau formalism, which requires only $4n^2 + O(n)$ classical bits to store~\cite{gidney2021stim}. 
An interesting aspect of quantum simulation circuits is that each building block, as shown in Figure~\ref{fig:HS_example}, typically consists of a single-qubit non-Clifford Rz rotation sandwiched between two Clifford subcircuits. This structure implies that many operations within these circuits can be efficiently simulated classically, which is crucial for our optimization strategies.

\subsection{Prior Works}
Because of its importance, many works have been proposed to optimize quantum simulation circuits. The prior works can be classified into three categories: simultaneous diagonalization~\cite{cowtan2019pytket,cowtan2020UCCSD_diagonalization, van2020diagonalization}, gate cancellation~\cite{li2021software_co_optimization, gui2020term,li2022paulihedral, jin2023tetris}, and Pauli network synthesis~\cite{de2024rustiq, paykin2023pcoast, schmitz2024graph_opt}. 

\textbf{Simultaneous Diagonalization} Early works on optimizing quantum simulation circuits are based on the simultaneous diagonalization of commuting Pauli strings. While many Pauli strings commute with each other, the Pauli strings can be partitioned into mutually commuting clusters, and the strings in each cluster can be simultaneously diagonalized to reduce the circuit cost and also maximize the cancellation between Pauli strings. The $\text{T}\vert \text{ket} \rangle$ compiler~\cite{sivarajah2020tket_compiler} leverages simultaneous diagonalization~\cite{cowtan2020UCCSD_diagonalization} and phase gadget synthesis~\cite{cowtan2019pytket} to optimize the circuits. 

\textbf{Gate Cancellation} The gates in Pauli string subcircuits can be cancelled between two similar Pauli strings. Li et al.~\cite{li2021software_co_optimization} propose a software and hardware co-optimization approach to synthesize circuits according to hardware constraints. The followup work Paulihedral~\cite{li2022paulihedral} generalizes gate cancellations and hardware-aware synthesis by proposing the Pauli intermediate representation (IR). Experimental results show that it outperforms simultaneous diagonalization-based methods. Jin et al. propose Tetris~\cite{jin2023tetris}, which introduces a refined IR that better considers gate cancellation and SWAP gate insertion when synthesizing for devices with limited connectivity.

\textbf{Pauli Network Synthesis}
The synthesis of quantum simulation circuits can be converted to the synthesis of a Pauli network, which is a generalization of the parity network~\cite{amy2018parity_network}.  Brugière et al. propose Rustiq~\cite{de2024rustiq}, an efficient approach for synthesizing quantum simulation circuits using Pauli networks. Unlike the symmetric structure in Paulihedral, where each CNOT tree is followed by an inverted CNOT tree to revert the transformations, Rustiq employs a bottom-up synthesis approach to find the smallest Clifford circuits to implement a Pauli string. 
Schmitz et al.~\cite{schmitz2024graph_opt} propose mapping a given Hamiltonian simulation circuit to a constrained path on a graph, referred to as the Pauli Frame Graph (PFG). 
Although interpreted differently, both works use Clifford circuits to transition from one Pauli string to another. The resulting circuit can also be viewed as a Pauli network.  Paykin et al. propose PCOAST~\cite{paykin2023pcoast}, a generalized compiler optimization framework based on the PFG approach, which has been integrated into the Intel Quantum SDK~\cite{wu2023intelsdk}. 

In our evaluation, we compare QuCLEAR with the state-of-the-art works from all three categories.

\section{Motivation}
\label{sec:motivation}
In this section, we discuss the key observations for our optimizations in the paper. Given that Clifford circuits can be simulated classically, a relevant question arises: Can we optimize quantum circuits by isolating the Clifford subcircuits and simulating them classically? Our optimizations leverage the unique properties of both Clifford circuits and quantum simulation circuits to achieve this goal.

\begin{figure}[htbp]
\centerline{\includegraphics[width=\linewidth]{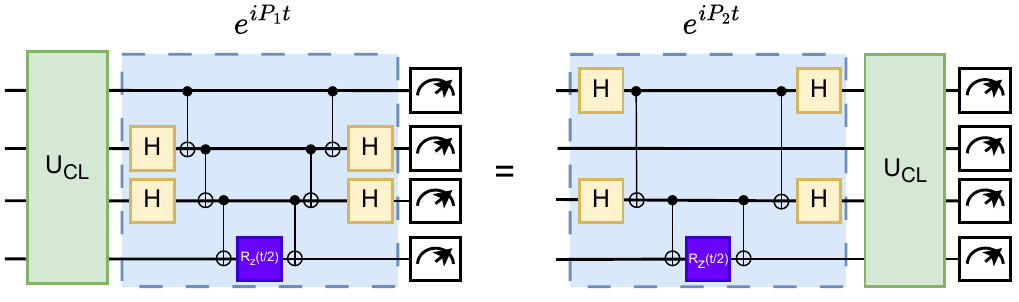}}
\caption{Clifford circuits weakly commute with Pauli rotation circuits, meaning that interchanging their positions transforms the Pauli string $P_1$ to another Pauli string $P_2$.}
\label{fig:CE_idea} 
\end{figure}

\textbf{Clifford Circuits and Quantum Simulation Circuits} An important property of quantum simulation circuits is that the circuit for implementing an exponentiated Pauli string (i.e., a Pauli rotation circuit) weakly commutes with Clifford gates~\cite{de2024rustiq}. To understand this weak commutation, consider that Clifford circuits stabilize the Pauli group. 
In other words, given a Pauli string $P_1$ and a Clifford circuit $U_{CL}$, the operation $U_{CL}^{\dagger} P_1 U_{CL}$ results in another Pauli string $P_2$. Therefore, for a Pauli rotation $e^{iPt}$, we have $e^{iP_1t} U_{CL} = U_{CL} e^{iP_2t}$, where $P_2 = U_{CL}^{\dagger} P_1 U_{CL}$. As shown in Figure~\ref{fig:CE_idea}, after switching the position of a Clifford circuit with a Pauli rotation, the resulting circuit is another Pauli rotation but with a different Pauli string $P_2$. Note that $P_2$ may include a negative sign. We omit this discussion here for simplicity. In our implementation, however, we track the sign, since it affects the sign of the rotation angle: $e^{i(-P)t} = e^{iP(-t)}$.
\begin{figure}[htbp]
\centerline{\includegraphics[width=\linewidth]{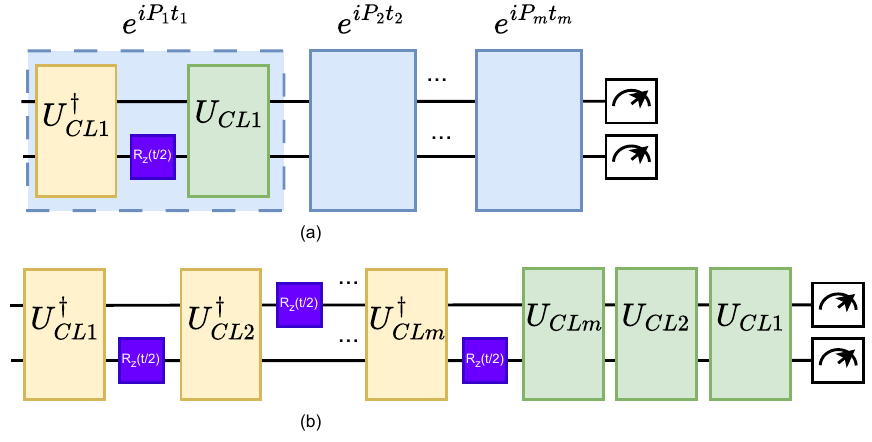}}
\caption{Extracting the Clifford subcircuit from a sequence of Pauli rotation blocks. (a) Identifying the best Clifford subcircuit design $U_{CL1}$ that maximizes the optimization of subsequent Pauli blocks. (b) Sequentially extracting Clifford subcircuits from each block to the end of the circuit.}
\label{fig:CE_demo} 
\end{figure}

This observation leads to our first optimization design, Clifford Extraction. We can select appropriate Clifford subcircuits and commute them with the Pauli rotations in a circuit. The extracted Clifford circuit can originate from a preceding Pauli rotation circuit. We show a simple optimization example in Figure~\ref{fig:opt_example}. The circuit represents $e^{iZZZZt_1}e^{YYXXt_2}$, and we measure the observable XXZZ. Since these two Pauli strings do not share common Pauli operators, the gate cancellation method in Paulihedral and Tetris can not optimize the circuit. As illustrated in Figures~\ref{fig:opt_example}(a) and (b), the Clifford subcircuit $U_{CL1}$ in $e^{iZZZZt_1}$ can be extracted to the end of the circuit. This extraction simplifies the second Pauli rotation circuit to $e^{iYYIIt_2}$, reducing the total CNOT gate count from 12 to 8. 

For a sequence of Pauli rotation blocks, we can sequentially extract the Clifford subcircuits from each block to the end of the circuit. As shown in Figure~\ref{fig:CE_demo}, we first determine the best Clifford subcircuit design $U_{CL1}$ that maximizes the optimization of subsequent Pauli blocks 
The extraction process is performed sequentially from the start to the end of the circuit, efficiently optimizing all Pauli blocks.  Clifford extraction also moves the Clifford subcircuits to the end, which is beneficial for subsequent Clifford Absorption optimization.

\begin{figure}[htbp]
\centerline{\includegraphics[width=\linewidth]{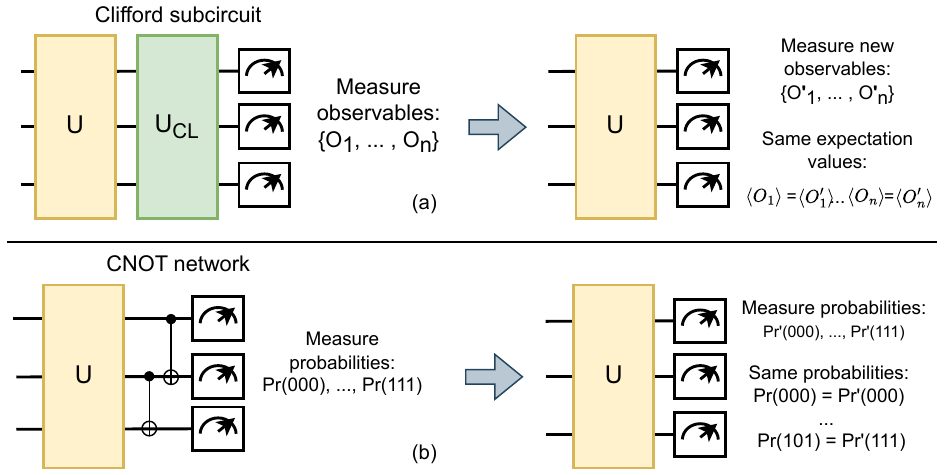}}
\caption{(a) The Clifford subcircuit at the end can be absorbed into the observable measurements. (b) The CNOT network at the end can be absorbed into the probability measurements.}
\label{fig:CA_idea} 
\end{figure}

\textbf{Clifford Circuits and Measurements} In this paper we propose two different approaches for absorbing Clifford subcircuits into measurements and classical post-processing. In many algorithms such as VQE for chemistry simulation, the focus is on measuring the expectation value of a set of Pauli observables ${O_{1},.., O_{n}}$, rather than measuring the probability distribution. Non-Pauli observables can be decomposed into a sequence of Pauli observables. For these observable measurements, we can optimize the circuit by absorbing the Clifford subcircuits into the measurement observables. Figure~\ref{fig:CA_idea}(a) illustrates the high-level idea.  Consider a circuit $U$ with Clifford subcircuit $U_{CL}$ at the end. Since Clifford circuits stabilize the Pauli group, the measured expectation value of an observable $O_1$ can be expressed as: $\bra{0} U^{\dagger} U^{\dagger}_{CL}O_1U_{CL} U \ket{0} = \bra{0} U^{\dagger} O'_1 U \ket{0}$. Thus, the result is equivalent to removing the Clifford subcircuit and measuring a new Pauli observable $O'_1 = U^{\dagger}_{CL}O_1U_{CL}$. In Figures~\ref{fig:opt_example}(b) and (c), the two extracted Clifford subcircuits at the end can be optimized while changing the observable from XXZZ to ZIXZ. By combining Clifford extraction and absorption, we reduce the total CNOT gate count from 12 to 4.

While chemistry simulation problems typically require measurements of observables, solving combinatorial optimization problems, such as using QAOA for MaxCut, requires measuring probabilities in the computational basis. For circuits that require probability distribution, we can consider a simplified case of Clifford subcircuits that consist solely of a CNOT network at the end of the circuit. The CNOT network is essentially a sequence of CNOT gates applied across multiple qubits. As we know, the CNOT gate transforms one computational basis state into another.  For example, $CNOT_{0,1}\ket{10} = \ket{10}$. By measuring the probability distribution before the CNOT network, we can efficiently calculate the output state after the CNOT network. This can also be explained by using the principle of deferred measurement~\cite{nielsen2010quantum}; in other words, measurement operations commutes with conditional operations. As we will demonstrate in Section~\ref{sec:CA}, the Clifford subcircuit in QAOA for combinatorial optimizations can be reduced to a single layer of Hadamard gates and a CNOT network.

Clifford extraction and Clifford absorption techniques complement each other effectively. In the next section, we will introduce the QuCLEAR framework, which integrates these two methods. We note that updating the Pauli strings and calculating the new observables involve multiplying the Clifford circuits with Pauli operators. These calculations can be performed efficiently on a classical computer by using the stabilizer tableau method~\cite{gottesman1998heisenberg, aaronson2004stabilizer_tableau}.

\begin{figure*}[htbp]
\centerline{\includegraphics[width=\linewidth]{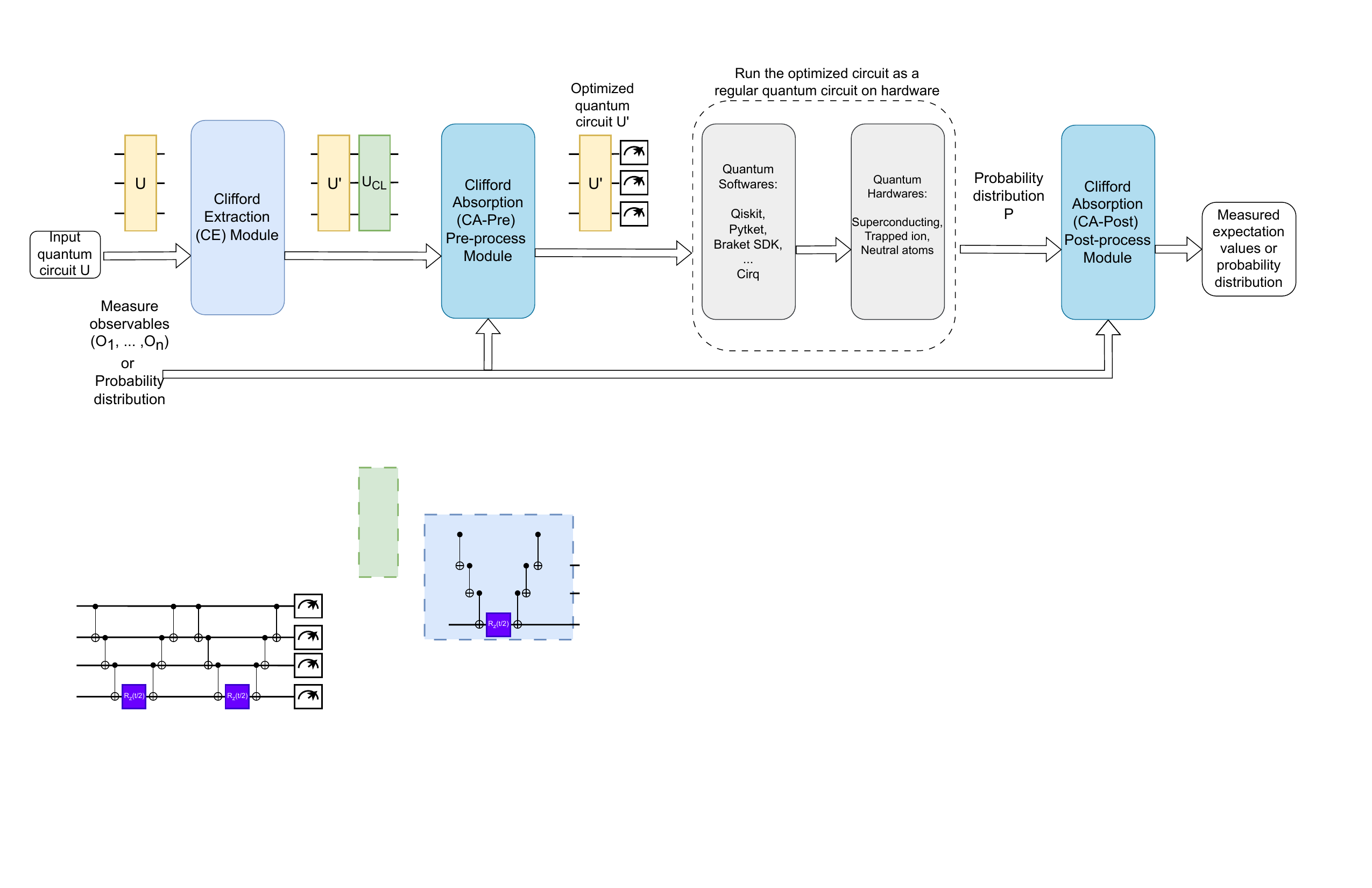}}
\caption{Overview of the QuCLEAR framework}
\label{fig:QuCLEAR_design} 
\end{figure*}

\section{QuCLEAR Framework}
\label{sec:framework}
In this section, we present a high-level overview of the QuCLEAR framework. 
Designed in a modular fashion to ensure portability and compatibility across different platforms, the QuCLEAR framework comprises three main components: the Clifford Extraction (CE) module, the Clifford Absorption pre-process (CA-Pre) module, and the Clifford Absorption post-process (CA-Post) module.

\begin{table*}[htbp]
\centering
\caption{New Pauli P' after commuting CNOT with original Pauli P. The left letter represents the control qubit, and the right letter represents the target qubit. The Pauli operators that have been reduced are marked in \textbf{bold}.}
\label{table:Paulis_CNOT}
\resizebox{\linewidth}{!}{%
\begin{tabular}{c||c|c|c|c|c|c|c|c|c|c|c|c|c|c|c|c|c|}
\hline
P & II & IX & IY & IZ & XI & XX & XY & XZ & YI & YX & YY & YZ & ZI & ZX & ZY & ZZ\\ \hline
P' & II & IX & ZY & ZZ & XX & \textbf{XI} & YZ & YY & YX & \textbf{YI} & XZ & XY & ZI & ZX & \textbf{IY} & \textbf{IZ}\\ \hline
\end{tabular}
}
\end{table*}

Figure~\ref{fig:QuCLEAR_design} shows the QuCLEAR workflow. The process begins with the Clifford Extraction (CE) module that optimizes the input circuit $U$ into an optimized circuit $U'$ followed by a Clifford subcircuit $U_{CL}$. This step preserves the unitary matrix of the original program, $U = U_{CL} U'$. Next, the optimized circuit $U'$ and the Clifford subcircuit $U_{CL}$ are passed to the Clifford Absorption pre-process (CA-Pre) module. The CA-Pre module also accepts the observables to be measured as input; if no observables are provided, it defaults to targeting the probability distribution. The CA-Pre module calculates the updated observables and appends the appropriate measurement basis to the end of the circuit. The optimized circuit $U'$ is then executed on quantum devices using any quantum software and hardware, since the framework is platform and software independent. After measuring the probability distribution from the quantum devices, the results are post-processed with the Clifford Absorption post-process (CA-Post) module. For experiments involving observable measurements, the CA-Post module provides the mapping between the original observables and the updated ones. For experiments focusing on probability measurements, the CA-Post module identifies the CNOT network at the end and calculates the mapping between the original probabilities and the measured probabilities.

\section{Clifford Extraction}
\label{sec:CE}
The Clifford extraction technique extracts Clifford subcircuits to the end of the circuit and optimizes each Pauli string it passes through. For a quantum simulation circuit $e^{iP_nt_n}...e^{iP_1t_1}$, we can sequentially extract the right Clifford subcircuits from each Pauli rotation subcircuit as shown in Fig.~\ref{fig:CE_demo}. Starting with the first Pauli $P_1$ and proceeding to the last Pauli $P_n$, this process effectively moves roughly half of the quantum simulation circuit's components to the end. Extracting Clifford subcircuits is not always advantageous, however, because the number of non-identity operators may increase after updating the Pauli strings. Previous work~\cite{zhang2019optimizing_extraction} has primarily focused on reducing T gates using this property. In this section, we propose a heuristic algorithm for identifying the Clifford subcircuit structure to best optimize the circuit.
\subsection{CNOT Tree Synthesis}
Various circuit structures are available for synthesizing a Pauli rotation $e^{iPt}$. As shown in Figure~\ref{fig:HS_example}, the Pauli rotation circuit consists of an Rz rotation gate sandwiched between mirrored Clifford subcircuits. The Clifford subcircuit includes a layer of single-qubit Clifford gates and a CNOT tree, where the root is the Rz rotation. The CNOT tree is essentially a mirrored version of the parity tree, with the root qubit encoding the parity of all the qubits with non-identity operator. The placements of the single-qubit Clifford gates are fixed, since they are determined by the Pauli string. However, the synthesis of the CNOT tree structure can vary. First, any qubit can serve as the root qubit. Second, once the root qubit is determined, the CNOT tree can be arranged to connect all the qubits associated with non-identity operators. While prior works propose CNOT tree or parity network synthesis for optimizing the quantum circuit, they primarily target more efficient logical or hardware-aware implementations of the CNOT tree or network~\cite{amy2018parity_network, nash2020quantum_parity,martiel2022architecture_parity, gheorghiu2022reducing_parity, de2022dynamic_parity, ender2023_parity, li2022paulihedral, jin2023tetris}. However, our approach focuses on selecting the CNOT tree that maximizes gate reduction post-commutation with Pauli blocks, yielding more substantial improvements than simply optimizing the tree structure locally.

\begin{figure*}[htbp]
\centerline{\includegraphics[width=\linewidth]{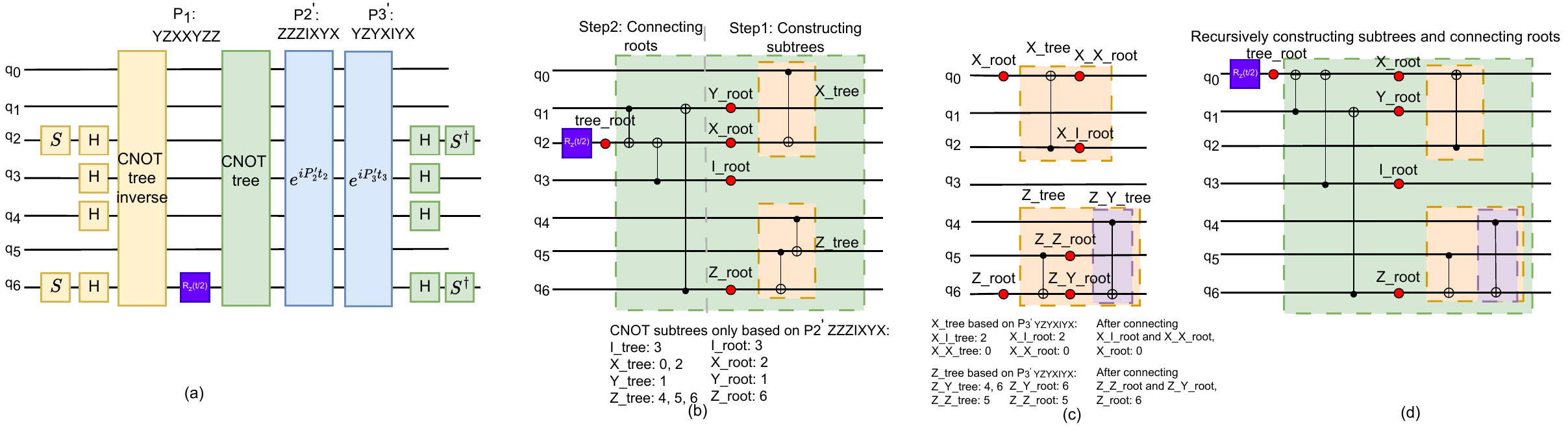}}
\caption{Synthesizing the CNOT tree circuit for P\textsubscript{1}. (a) Circuit after extracting the single-qubit gates in $e^{iP_1t}$ (b) Detailed view of tree circuit synthesis for optimizing P\textsubscript{2}'. (c) Synthesis process for the two subtrees, X\_tree and Z\_tree based on P\textsubscript{3}'. (d) Synthesized tree circuit for optimizing P\textsubscript{2}' and P\textsubscript{3}'}
\label{fig:tree_syn_example} 
\end{figure*}

We first consider a simplified case, focusing on commuting a CNOT gate with a two-qubit Pauli string. The original Pauli P and the new Pauli P' after commuting are listed in Table~\ref{table:Paulis_CNOT}. As shown in the table, the four Pauli strings XX, YX, ZY, and ZZ can be optimized. The Pauli Z on the control qubit and the Pauli X on the target qubit might be optimized to I. XX and ZZ are special cases where the pattern can be extended to multiple qubits. An n-qubit Pauli string with all Z operators ZZ...ZZ, can be optimized by commuting with a chain of CNOT gates to generate a Pauli string with only one non-identity operator:  II...IZ. Similarly, an n-qubit Pauli string with all X operators XXXX...X, can also be optimized by commuting with a chain of CNOT gates to generate a Pauli with $\left\lceil \frac{n}{2} \right\rceil$ non-identity operators: XIXI...X. An n-qubit Pauli string with all Y operators YYYY...Y, can be commuted with a chain of CNOT gates to generate a Pauli with $\left\lceil \frac{n}{2} \right\rceil$ non-identity operators: XIXI...Y. An n-qubit identity Pauli string II...I remains an identity after commuting with CNOT gates. The observations lead to our first optimization strategy: grouping identical Pauli operators together.

We will use an example in Figure~\ref{fig:tree_syn_example} to demonstrate the process of synthesizing the CNOT tree. The circuit contains three Paulis: P\textsubscript{1}: YZXXYZZ, P\textsubscript{2}: YZXIZYX, and P\textsubscript{3}: XZYZIYX. Our objective is to find the optimal CNOT tree structure for P\textsubscript{1}, which can optimize P\textsubscript{2} and P\textsubscript{3}. 
Since the single-qubit gates are not changeable, we will extract the single-qubit gates and update Paulis to P\textsubscript{2}': ZZZIXYX and P\textsubscript{3}': YZYXIYX. The circuit after extracting the gates is shown in Figure~\ref{fig:tree_syn_example}(a).

We will find the best CNOT tree structure for optimizing the immediately following Pauli P\textsubscript{2}. This greedy algorithm is based on the understanding that Pauli operators appearing later in the sequence have more optimization opportunities, because they can be optimized with multiple Clifford subcircuits. Therefore, we prioritize optimizing the first succeeding Pauli string. Based on the observation that identical Pauli operators should be grouped together and processed through a CNOT tree, we first partition the Pauli operators into four groups and synthesize four corresponding subtrees: I\_tree, X\_tree, Y\_tree, and Z\_tree. A subtree in a parity tree is a smaller section of the tree, where the root of the subtree encodes the parity of a specific subset of qubits. For example, I\_tree denotes the CNOT parity tree that only encodes the parity of the qubits associated with I operators. The complete CNOT tree that encodes the parity of all qubits can be constructed by using CNOT gates to connect roots of all four subtrees. When optimizing solely for P\textsubscript{2}', the order in which qubits are connected within a subtree is unimportant, since their corresponding Pauli operators in P\textsubscript{2}' are identical. Figure~\ref{fig:tree_syn_example}(b) provides a detailed view of the CNOT tree synthesis. We sequentially connect the qubits within each subtree, designating the last qubit as the root of that subtree. After synthesizing the subtrees, we have four root qubits that need to be connected. According to Table~\ref{table:Paulis_CNOT}, we observe that combinations YX and ZY can be optimized, while IX remains unchanged. Therefore we will prioritize connecting the Z\_root with Y\_root, I\_root with X\_root, and Y\_root with X\_root. The final root qubit is designated as the root for the CNOT tree, where we will place the Rz rotation gate. In our example in Figure~\ref{fig:tree_syn_example}(b), after extracting the synthesized CNOT tree, P\textsubscript{2}': ZZZIXYX is optimized into IIIIXYX. 

While the order in which qubits are connected within a subtree is irrelevant for optimizing P\textsubscript{2}', it may become important for subsequent Pauli strings. This consideration leads to the discussion of the recursive algorithm in the next subsection.

\subsection{Recursive CNOT Tree Synthesis}
The CNOT tree can be synthesized by recursively generating the subtrees. When optimizing P\textsubscript{2}', the two subtrees X\_tree and Z\_tree have the freedom of choosing their root qubit and the order of connecting the qubits with the CNOT gates. To find the optimal gate order, we consider the next Pauli P\textsubscript{3}': YZYXIYX. For synthesizing the subtree Z\_tree, the only requirement is to connect the three qubits q\textsubscript{4}, q\textsubscript{5}, and q\textsubscript{6}. Thus, we examine the corresponding three Pauli operators in P\textsubscript{3}' that are YZY. Figure~\ref{fig:tree_syn_example}(c) details the synthesis of the two subtrees X\_tree and the Z\_tree. For Z\_tree, we follow the same process: synthesizing subtrees Z\_Y\_tree and Z\_Z\_tree. Then we connect the root qubits Z\_Y\_root and Z\_Z\_root using the previously discussed root connection strategy. Here, Z\_Y\_tree refers to the subtree that connects the qubits associated with Z in P\textsubscript{2}' and Y in P\textsubscript{3}'. This process is equivalent to calling the tree synthesis algorithm with a smaller Pauli substring of P\textsubscript{3}': P\textsubscript{3}'[4:7]. Therefore, we can recursively call the tree synthesis algorithm to build subtrees and connect the root qubits. The final circuit for optimizing both P\textsubscript{2}' and P\textsubscript{3}' is shown in Figure~\ref{fig:tree_syn_example}(d). If we extract the CNOT tree in Figure~\ref{fig:tree_syn_example}(b), P\textsubscript{3}' YZYXIYX will be converted into IXXXXYX where the number of non-identity operators is not reduced. However, if we extract the optimized CNOT tree in Figure~\ref{fig:tree_syn_example}(d),  P\textsubscript{3}' will be optimized into IIXXIYX, converting two non-identity operators into identity operators.

\begin{algorithm}
\caption{tree\_synthesis}
\label{alg:Tree_synthesis}
\begin{algorithmic}[1]
\REQUIRE Pauli\_strings $P$, Pauli\_index $p\_idx$, Quantum circuit $qc$, Tree indexes $tree\_idxs$, extracted Clifford circuit \textit{extr\_clf}.
\ENSURE Append a synthesized CNOT tree to quantum circuit $qc$ based on the Pauli strings in $P$,  and return the tree root index $tree\_root$.
\STATE $I\_tree = X\_tree = Y\_tree = Z\_tree = []$
\STATE $next\_idx = p\_idx + 1$
\STATE $next\_pauli =$ update\_pauli($P[next\_idx]$, \textit{extr\_clf})
\FOR{each $index$ in $tree\_idxs$}
    \STATE \textbf{switch} $next\_pauli[index]$ \textbf{do}
    \STATE \quad \textbf{case} $X$, $Y$, $Z$, $I$: Append $index$ to $X\_tree$, $Y\_tree$, $Z\_tree$ or $I\_tree$
    \STATE \textbf{end switch}
\ENDFOR

\FOR{each $subtree$ in $\{X\_tree, Y\_tree, Z\_tree, I\_tree\}$}
    \IF{$\text{len($subtree$)} = 1$}
        \STATE $subtree\_root \leftarrow subtree[0]$
    \ELSIF{$\text{len($subtree$)} > 1$}
        \STATE $subtree\_root \leftarrow $ tree\_synthesis($P$, $next\_idx$, $qc$, $subtree$, \textit{extr\_clf})
    \ENDIF
    \STATE Assign $subtree\_root$ to the corresponding root variable ($X\_root, Y\_root, Z\_root, I\_root$)
\ENDFOR

\STATE $all\_roots$ = $(Z\_root, I\_root, Y\_root, X\_root)$
\FOR{each $root\_name$ in $all\_roots$}
    \STATE $tree\_root \leftarrow \text{connect\_roots}$ ($qc$, $root\_name$, $all\_roots$)
\ENDFOR

\RETURN $tree\_root$

\end{algorithmic}
\end{algorithm}

The recursive algorithm for synthesizing a CNOT tree is outlined in Algorithm~\ref{alg:Tree_synthesis}. The input is the Pauli string for synthesis. The \textit{p\_idx} is the index of the Pauli to be synthesized. The \textit{qc} is the quantum circuit to add CNOT gates. The \textit{tree\_idxs} is the list of indexes of the current tree. \textit{extr\_clf} is the already extracted Clifford circuit, which will be used to calculate the updated Pauli $U^{\dagger}_{CL}PU_{CL}$. After initializing all the variables, we will find the next Pauli string and use the \textit{update\_pauli} function to calculate the updated next Pauli. For every qubit index in \textit{tree\_idxs}, we examine the corresponding Pauli operator in the next Pauli string \textit{next\_pauli} and append the qubit index to the appropriate subtree. Then we evaluate each subtree. If a subtree contains only one qubit, we designate it as the \textit{subtree\_root}. Otherwise, we need to determine the order of the CNOT gates for that subtree. We recursively call the \textit{tree\_synthesis} algorithm to construct the subtree and return the \textit{subtree\_root}. Once all the subtrees are generated, we connect the roots in sequence. The function \textit{connect\_roots} is invoked to add CNOT gates to the quantum circuit $qc$ between the root qubits of the subtrees. The last root becomes the return value of the \textit{tree\_synthesis} algorithm.

\subsection{Clifford Extraction Algorithm}

\begin{algorithm}
\caption{Clifford Extraction Algorithm}
\label{alg:CE_algorithm}
\begin{algorithmic}[1]
\REQUIRE List of Pauli strings $P$, 
\ENSURE Synthesize an optimized quantum circuit $opt\_qc$ and the extracted Clifford circuit \textit{extr\_clf}.
\STATE $commute\_blocks$ = convert\_commute\_sets($P$)
\FOR{$block$ in $commute\_blocks$}
    \FOR{$pauli\_idx$ in range(len($block$))}
        \STATE $curr\_pauli = block[pauli\_idx]$
        \STATE $next\_idx$ = find\_next\_pauli($curr\_pauli$, $pauli\_idx$, $block$, \textit{extr\_clf})
        \STATE $next\_pauli$ = $block.pop(next\_idx)$
        \STATE $block.insert(pauli\_idx + 1, next\_pauli)$
        \STATE $cx\_tree$ = QuantumCircuit()
        \STATE $tree\_root$ = tree\_synthesis($block$, $pauli\_idx$, $cx\_tree$, \textit{extr\_clf})
        \STATE Update $opt\_qc$ and \textit{extr\_clf} based on the CNOT tree $cx\_tree$, the root qubit $tree\_root$, and $curr\_pauli$.
    \ENDFOR
\ENDFOR
\STATE \textbf{return} $opt\_qc$, \textit{extr\_clf}
\end{algorithmic}
\end{algorithm}

In this subsection we present our Clifford Extraction algorithm. The pseudo code is listed in Algorithm~\ref{alg:CE_algorithm}; for simplicity we omit the parameters associated with the Paulis. The algorithm takes a list of Pauli strings as input and outputs the optimized quantum circuit \textit{opt\_qc}, and the extracted Clifford circuit \textit{extr\_clf}. The algorithm begins with converting the Pauli list $P$ into blocks of commuting Paulis. These blocks capture local commuting information, allowing Paulis within each block to switch order. This reordering is useful for optimizing the circuit. The blocks can be constructed using a simple algorithm that compares each subsequent Pauli operator with those in the current block. If all commute, the Pauli is added to the block; otherwise, a new block is created. This conversion is scalable, with a complexity of $O(nm^2)$, where n and m are the number of qubits and Pauli strings. We note that this approach differs from the block definition in Paulihedral. In QuCLEAR, the Paulis within a block can change order, but the order of the blocks themselves remains fixed. In contrast, Paulihedral allows the block order to change, enabling more global optimization. This difference in design choices stems from our decision not to assume any high-level information or prior knowledge about the benchmarks. 

Since Paulis commute within a block, QuCLEAR uses the \textit{find\_next\_pauli} function to identify the most suitable Pauli in the current commuting block. The \textit{find\_next\_pauli} function implements a simple greedy algorithm that traverses the current block, computes the optimized Pauli for each Pauli string, and selects the Pauli with the fewest non-identity operators after extracting the Clifford circuit from the current Pauli string. The cost associated with each Pauli is determined by running \textit{tree\_synthesis} without recursion to compute the optimized Pauli string. After finding the next Pauli, we use the \textit{tree\_synthesis} function to recursively generate the CNOT tree and return the root qubit index \textit{tree\_root}. Then we update the optimized circuit \textit{opt\_qc} and the already extracted Clifford circuit \textit{extr\_clf}. After traversing all Paulis in the list $P$, we generate the optimized quantum circuit and the extracted Clifford subcircuit \textit{extr\_clf}.

During the execution of the algorithm, the \textit{extr\_clf} variable keeps track of the already extracted Clifford circuits. This variable is used to update the next Pauli string. Whenever we determine the CNOT tree design, we defer the extraction of the CNOT tree to avoid the overhead of updating all subsequent Pauli strings. Instead, we update the Pauli string only when needed, specifically when it becomes the \textit{next\_pauli} in the CNOT tree synthesis process.

\subsection{Scaling}
The Clifford Extraction algorithm is scalable with respect to both the number of qubits $n$ and the number of Pauli strings $m$.  First, converting Paulis into commuting blocks requires comparing each Pauli with every other Pauli and checking each bit, resulting in a worst-case complexity of $O(nm^2)$ operations. Subsequently, for each Pauli, the algorithm searches for the optimal next Pauli within the current commuting block. In the worst-case scenario, where all Paulis commute, the \textit{tree\_synthesis} function needs to be called $O(m^2)$ times. For each \textit{tree\_synthesis} run, the maximum recursive depth is $O(n)$, and the most time consuming step in each recursive call is to update the Pauli operators with the already extracted Cliffords. Based on the stabilizer tableau~\cite{gottesman1998heisenberg, aaronson2004stabilizer_tableau}, the update of the Clifford operators requires only $O(n^2)$ operations on classical computers. Therefore, the worst-case time complexity of \textit{tree\_synthesis} is $O(n^3)$, leading to an overall worst-case complexity of $O(n^3m^2)$. The algorithm shares the same complexity as the Rustiq algorithm~\cite{de2024rustiq}. However, in the best-case scenario, where none of the Paulis commutes, the time complexity is reduced to $O(n^3m)$. As demonstrated in our experiments in Section~\ref{sec:evaluation}, our approach achieves significantly shorter compile times when the order of the Pauli operators remains relatively unchanged.


\begin{figure*}[htbp]
\centerline{\includegraphics[width=\linewidth]{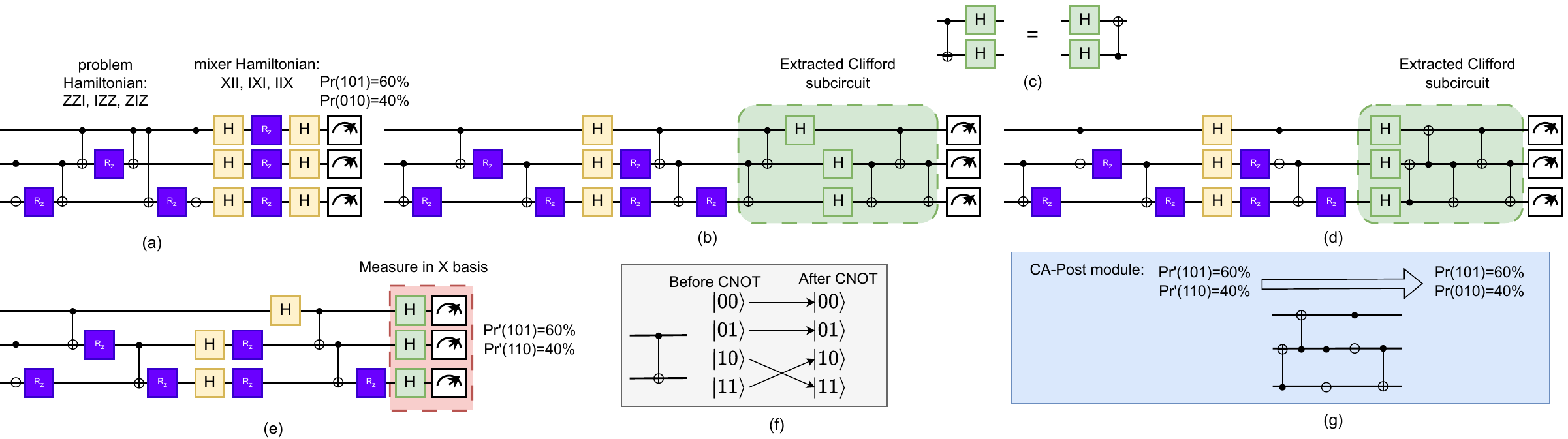}}
\caption{Optimizing the QAOA circuit with Clifford Absorption. (a) Original QAOA circuit. (b) Optimized circuit after Clifford Extraction. (c) Layer of Hadamard gates that commutes with CNOT gate. (d) The circuit after commuting the Hadamard gates. (e)Optimized circuit after absorbing the CNOT network. (f) A CNOT gate performs an XOR operation on the basis states. (g) The CA-Post module updates the distribution based on the measured probabilities and the CNOT network.}
\label{fig:QAOA_opt} 
\end{figure*}

\section{Clifford Absorption}
\label{sec:CA}
The Clifford Absorption modules behave slightly differently depending on whether observables or probability distributions are being measured. We will discuss each scenario in the following subsections.
\subsection{Observable Measurement}
When applying quantum simulation circuits to simulate chemical systems, the primary research interest is in measuring observables. These observables are typically represented as a set of Pauli strings $O_1$,...,$O_m$. For example, an n-qubit VQE algorithm requires measuring $O(n^4)$ Pauli observables for molecular Hamiltonians~\cite{wecker2014N4measurement, hastings2014N4measurement}. As discussed in Section~\ref{sec:motivation}, the Clifford subcircuit $U_{CL}$ extracted by the Clifford Extraction module can be absorbed into these observables. The new observables are $O_i' = U^{\dagger}_{CL}O_i U_{CL}$ which are also Pauli observables. Prior work have leveraged the property to optimize variational algorithms, either by appending Clifford circuits to expand the search space~\cite{shang2023schrodinger} or error mitigation~\cite{seifert2024clapton_error_mitigation}. Yet, it has not been widely adopted as a generalized compiler optimization, as the number of Clifford gates at the end of the circuit is typically very limited.  Our work is the first to demonstrate how Clifford Extraction and Absorption can have a significant effect in reducing circuit size.

To measure these new Pauli observables, we adjust the measurement basis by adding single-qubit rotations at the end of the circuit~\cite{nielsen2010quantum}. The CA-Pre module calculates the new observables using the stabilizer tableau and pre-processes the circuits by adding the corresponding single-qubit gates. While our default implementation measures these observables separately (requiring one circuit per observable), it does not preclude the possibility of further optimizations. Because of the property of Clifford operations preserving the commutation and anti-commutation relationships between Pauli operators, the transformed observables retain these relationships. We can apply the commutation-based measurement reduction technique~\cite{gokhale2020n3measurement} to reduce the number of measurements to $O(n^3)$ or use the classical shadow~\cite{nakaji2023classical_shadow_measurement} to estimate the expectation value of the observables. The CA-Post module for observable measurement functions as a dictionary, mapping the measurement results to the correct observables.

The CA modules are highly scalable with respect to the number of qubits. The update of the Clifford operators requires only $O(n^2)$ operations on classical computers. Additionally, it takes $4n^2 + O(n)$ classical bits~\cite{gidney2021stim} to store the stabilizer tableau. To measure $k$ observables, the total classical computation overhead of the CA modules is $O(kn^2)$. 

\subsection{Probability Distribution Measurement}
When solving the combinatorial optimization problem using QAOA, the result is measured in the computational basis, and the classical bitstring corresponds to the solution of the problem.

We propose a scaling approach based on the properties of the QAOA algorithm. In the QAOA, two primary Hamiltonians are employed iteratively: the problem Hamiltonian (also known as the cost Hamiltonian) and the mixer Hamiltonian. These Hamiltonians play distinct roles in encoding the optimization problem and exploring the solution space. Combinatorial optimization problems are encoded in the problem Hamiltonian, which typically consists only of Pauli Z and I operators, as they represent classical terms. For instance, Quadratic Unconstrained Binary Optimization (QUBO)~\cite{lucas2014isingQUBO} problems can be expressed as a sequence of Pauli Z and ZZ rotations in QAOA. Other combinatorial optimization problems, such as Low Autocorrelation Binary Sequences (LABS)~\cite{boehmer1967binaryLABS, shaydulin2024evidenceLABS}, might require more multi-qubit Pauli Z rotations. The mixer Hamiltonian ensures that the quantum state does not get trapped in local minima and can explore different configurations. It typically consists of $n$ Pauli strings with a Pauli X rotation acting on each qubit. 

Figure~\ref{fig:QAOA_opt}(a) illustrates a single-layer QAOA circuit for the MaxCut problem. The problem Hamiltonian is $ZZI + IZZ + ZIZ$, and the mixer Hamiltonian is $XII + IXI + IIX$. We can extract the Clifford subcircuit using the Clifford Extraction technique. The resulting circuit is shown in Figure~\ref{fig:QAOA_opt}(b). Since the original Hamiltonians contain only Pauli I, X, and Z operators, the extracted Clifford circuit consists only of Hadamard gates and CNOT gates. By leveraging the circuit equivalence shown in Figure~\ref{fig:QAOA_opt}(c), we can commute the Hadamard gates with the CNOT gates. The Clifford subcircuit is equivalent to a single layer of Hadamard gates and a CNOT network, which can be classically simulated. The final optimized circuit is shown in Figure~\ref{fig:QAOA_opt}(e), with CNOT gate count reduced from 6 to 5. While the reduction of just one CNOT gate may seem minor, the previous circuit was already highly optimized, and any improvement is valuable. We will demonstrate in Section~\ref{sec:evaluation} that for larger circuits and more complicated problems, the reduction in gate count becomes more significant. Figure~\ref{fig:QAOA_opt}(f) and (g) show the process of classically updating the probability distribution. A CNOT gate applies an XOR operation on the basis states. In the optimized circuit, the measured probabilities are $Pr'(101)=60\%$ and $Pr'(110)=40\%$. The CA-Post module then computes the resulting states after the CNOT network, yielding updated probabilities of $Pr(101)=60\%$ and $Pr(010)=40\%$. When resolving the CNOT network classically, the complexity depends on the number of shots used to collect the probability distribution. 
We will provide a detailed scalability analysis in the following paragraph.

In this section we prove that the Clifford subcircuit extracted in the QAOA algorithm can be optimized to a single layer of Hadamard gates followed by a CNOT network.
\begin{prop}
In the QAOA algorithm, for any problem Hamiltonian consisting only of Pauli I and Z operators and a mixer composed of Pauli X rotations, the extracted Clifford subcircuit can be reduced to a single layer of Hadamard gates and a CNOT network. 
\end{prop} 

To establish this result, we define a sequence of Pauli strings consisting solely of Pauli Z and I operators as a Z-I Pauli sequence.  The problem Hamiltonian in QAOA consists of Z-I Pauli sequences, while the mixer Hamiltonian consists of X-I Pauli sequences. We then present the following lemma.

\begin{lemma} Commuting a CNOT network with a Z-I Pauli sequence results in a new Z-I Pauli sequence. Commuting a CNOT network with an X-I Pauli sequence results in a new X-I Pauli sequence.
\label{lem:lemma1}
\end{lemma}

This can be straightforwardly demonstrated by noting that a single CNOT gate transforms a Pauli string containing Z and I operators into another Pauli string of the same type, and the same logic applies to X-I Pauli sequences. Based on Lemma~\ref{lem:lemma1}, the Clifford subcircuit extracted from a Z-I Pauli sequence is a CNOT network since each Z-I Pauli string, when commuted with the CNOT network, results in another Z-I Pauli string. The extracted Clifford circuit will contain only CNOT gates.

The mixer Hamiltonian consists of an X-I Pauli sequence. Commuting a CNOT network with the mixer Hamiltonian generates a new X-I Pauli sequence. When performing Clifford extraction for the updated mixer layer, we extract a CNOT network and a layer of Hadamard gates. The extracted Hadamard layer converts the next Z-I Pauli sequence into an X-I Pauli sequence, and the CNOT network subsequently transforms the X-I Pauli sequence into another X-I Pauli sequence. Therefore, for a QAOA algorithm with $j$ layers of problem and mixer Hamiltonians, the extracted Clifford subcircuit consists of $2j-1$ layers of Hadamard gates and $2j$ layers of CNOT networks. As shown in Figure~\ref{fig:QAOA_opt}(d), the extracted circuit consists of one layer of Hadamard gates and followed by a CNOT network at the end. 

When measuring probability distributions for QAOA, the CA-Pre module adds a single layer of Hadamard gates at the end of the optimized circuit. The CA-Post module will calculate the updated probability distribution based on the measured probabilities and the CNOT network. The calculation of the CA-Post module is also scalable. As shown in Figure~\ref{fig:QAOA_opt}(f)(g), the CA-Post module will update the probability distribution by performing XOR operations on the measured states based on each CNOT gate. For a QAOA problem with $k$ shots and measured $s$ states with non-zero probability, only these $s$ states need to be updated, rather than the total $2^n$ states; $s$ should always be smaller than or equal to the number of shots $k$. In the worst case, where each measured state corresponds to a unique shot, and there are $m$ CNOTs in the CNOT network, the classical computation overhead of the CA module is $O(mk)$. In practice, the number of shots is typically a constant number, such as 1000.

\section{Methodology}
\label{sec:methodology}

\textbf{Benchmarks:} We evaluate 19 benchmarks in our experiments, including different sizes and applications. For the chemistry eigenvalue problem, we select the UCCSD ansatz using the Jordan-Wigner transformation~\cite{barkoutsos2018UCCSD_JW} with different numbers of electrons and orbitals. For instance, UCC-(2,4) refers to a configuration with two electrons and four spin orbitals, representing a valid active space for simulating H$_2$. We also include the Hamiltonian simulation for LiH, H\textsubscript{2}O, and the benzene molecule, with well-established active spaces~\cite{ryabinkin2018qubit, smart2022resolving}. For the combinatorial optimization problems, we select the QAOA algorithm for solving MaxCut problems on regular graphs of degrees 4, 8, and 12 and on random graphs. MaxCut-(n15, r4) refers to a regular graph with 15 nodes, each having a degree of 4. MaxCut-(n15, e63) refers to a random graph with 15 nodes and 63 edges. Additionally, we evaluate the QAOA for solving LABS problems, which have recently demonstrated a scaling advantage over classical algorithms~\cite{shaydulin2024evidenceLABS}. The QAOA for a LABS problem contains multi-qubit Pauli Z rotations in the problem Hamiltonian. All the QAOA benchmarks contain one iteration of the problem and mixer Hamiltonian. The benchmarks and the native number of gates, without any optimizations, are listed in Table~\ref{table:benchmarks}.

\begin{table}[htbp]
\centering
\caption{Benchmark information}
\label{table:benchmarks}
\resizebox{\linewidth}{!}{%
\begin{threeparttable}
\begin{tabular}{|c|c|c|c|c|c|}
\hline
Type & Name & \#qubits & \#Pauli & \#CNOT & \#1Q \\ \hline
\multirow{6}{*}{UCCSD} & UCC-(2,4) & 4 & 24 & 128 & 264\\ \cline{2-6}
 & UCC-(2,6) & 6 & 80 & 544 & 944\\ \cline{2-6}
 & UCC-(4,8) & 8 & 320 & 2624 & 3968\\ \cline{2-6}
 & UCC-(6,12) & 12 & 1656 & 18048 & 21096\\ \cline{2-6}
& UCC-(8,16) & 16 & 5376 & 72960 & 69120\\ \cline{2-6}
& UCC-(10,20) & 20 & 13400 & 217600 & 173000\\ \hline
\multirow{3}{*}{\makecell{Hamiltonian \\ simulation}} & LiH &   6 &61 & 254 & 421\\ \cline{2-6}
 & H\textsubscript{2}O & 8  & 184 & 1088 & 1624\\ \cline{2-6}
  & benzene & 12 & 1254 & 10060 & 12390\\ \hline

\multirow{3}{*}{\makecell{QAOA \\ LABS}} & LABS-(n10) & 10 & 80 & 340 & 100\\ \cline{2-6}
 & LABS-(n15) & 15 & 267 & 1316 & 297 \\ \cline{2-6}
  & LABS-(n20) & 20 & 635 & 3330 & 675\\ \hline
\multirow{6}{*}{\makecell{QAOA \\ MaxCut}} & MaxCut-(n15, r4) & 15 & 45 & 60 & 75\\ \cline{2-6}
 & MaxCut-(n20, r4) & 20 & 60 & 80 & 100 \\ \cline{2-6}
 & MaxCut-(n20, r8) & 20 & 100 & 160 & 140\\ \cline{2-6}
 & MaxCut-(n20, r12) & 20 & 140 & 240 & 180\\ \cline{2-6}
 & MaxCut-(n10, e12) & 10 & 22 & 24 & 42\\ \cline{2-6}
 & MaxCut-(n15, e63) & 15 & 78 & 126 & 108\\ \cline{2-6}
  & MaxCut-(n20, e117) & 20 & 137 & 234 & 177\\ \hline
  
\end{tabular}
\begin{tablenotes}
    \normalsize
    \item \#1Q denotes the number of single-qubit operations
    \end{tablenotes}
\end{threeparttable}
}
\end{table}

\textbf{Implementation:} We prototype our QuCLEAR framework in Python3.10 on top of the quantum computing framework Qiskit~\cite{qiskit2024}. The implementation is open-sourced~\cite{QuCLEAR}.

\textbf{Comparison with State-of-the-Art Methods} : We compared QuCLEAR with Qiskit version 1.1.1~\cite{qiskit2024}, $\text{T}\vert \text{ket} \rangle$ version 1.30.0~\cite{cowtan2019pytket}, Paulihedral~\cite{li2022paulihedral}, Rustiq~\cite{de2024rustiq}, and Tetris~\cite{jin2023tetris}. The baselines encompass a range of methods. $\text{T}\vert \text{ket} \rangle$ employs the simultaneous diagonalization method~\cite{cowtan2019pytket, cowtan2020UCCSD_diagonalization}. Paulihedral focuses on gate cancellation. Rustiq employs Pauli network synthesis. 
All the baselines have been updated to the latest versions available at the time of writing this paper. QuCLEAR, Paulihedral, and Tetris are optimized with Qiskit optimization level 3. $\text{T}\vert \text{ket} \rangle$ is optimized with its own optimization pass with maximum optimization setting O2. We did not optimize $\text{T}\vert \text{ket} \rangle$ circuits with Qiskit optimization passes because previous research~\cite{jin2023tetris} indicates that $\text{T}\vert \text{ket} \rangle$ performs best with its own optimizer. Rustiq is not further optimized because the Rustiq-optimized circuit outputs only the Pauli network and omits the single-qubit Rz rotation gates. We choose the $rcount$ implementation in Rustiq to minimize the CNOT gate counts. We did not compare with QAOA-specific compilers, as Paulihedral and Tetris have been shown to outperform~\cite{alam2020circuit_QAOA_PH} and \cite{lao20222qan_QAOA_Tetris}, respectively. We also conducted experiments on two backends with limited connectivity: the 65-qubit IBM\_Manhattan with a heavy-hex lattice architecture, and the 64-qubit Google Sycamore with a 2D grid architecture.

We compared the CNOT gate count, the entangling depth (also known as CNOT-depth), and the compile time across different methods. We compared the entangling depth instead of the circuit depth to ensure a fair comparison, since Rustiq's results do not include the single-qubit rotation gates.
The evaluations were performed on a Windows 10 device with an Intel Core i7-10870H CPU @ 2.20 GHz, NVIDIA GeForce RTX 3080 Laptop GPU, 32.0 GB RAM, and 1 TB SSD.


  

\begin{table*}[htbp]
\centering
\caption{Results and compile time compared with the state-of-the-art works on a fully connected device}
\label{table:FC_results}
\resizebox{\linewidth}{!}{%
\begin{threeparttable}
\begin{tabular}{|c|c|c|c|c|c|c|c|c|c|c|c|c|c|c|c|}
\hline
 \multirow{2}{*}{Name}  & \multicolumn{5}{|c|}{\#CNOT} & \multicolumn{5}{|c|}{Entangling depth}& \multicolumn{5}{|c|}{Compile time (s)} \\ \cline{2-16}
 &QuCLEAR &Qiskit &Rustiq & PH &tket &QuCLEAR &Qiskit &Rustiq &PH &tket &QuCLEAR &Qiskit &Rustiq &PH & tket\\ \hline
 UCC-(2,4) & \textbf{23} & 41 & 33 & 48 & 53  & \textbf{17} & 41 & 31 & 48 & 50  & 0.1327 & 0.1546 & \textbf{0.001} & 0.1356 & 2.2191\\ \hline
 UCC-(2,6) & \textbf{106} & 181 & 161 & 216 & 236  & \textbf{82} & 178 & 145 & 215 & 211  & 0.561 & 0.5823 & \textbf{0.007} & 0.5864 & 9.2433  \\ \hline
 UCC-(4,8) & \textbf{448} & 1003 & 795 & 947 & 1257  & \textbf{335} & 990 & 666 & 941 & 1112  & 3.0018 & 2.5905 & \textbf{0.0848} & 3.2474 & 49.1337\\\hline
 UCC-(6,12) & \textbf{2580} & 5723 & 4705 & 6076 & 8853  & \textbf{1832} & 5582 & 3588 & 6051 & 7924  & 21.656 & 19.6461 & \textbf{10.7976} & 24.1745 & 365.9686 \\ \hline
 UCC-(8,16) & \textbf{8820} & 23498 & 18552 & 23389 & 36369  & \textbf{6153} & 23020 & 13092 & 23333 & 33074  & 98.4158 & \textbf{96.1222} & 759.8567 & 137.9728 & 2224.6543 \\ \hline
UCC-(10,20) & \textbf{24302} & 72005 & 50913 & 67336 & 109130  & \textbf{15979} & 70705 & 34403 & 67237 & 100585  & \textbf{306.3307} & 342.6219 & 14802.1183 & 629.8874 & 14278.8268 \\ \hline
 LiH & \textbf{74} & 180 & 114 & 121 & 132  & \textbf{60} & 176 & 92 & 119 & 112  & 0.3999 & 0.3701 & \textbf{0.0059} & 0.3939 & 5.0096  \\ \hline
 H\textsubscript{2}O & \textbf{274} & 786 & 350 & 471 & 505  & \textbf{189} & 769 & 267 & 461 & 442  & 1.7922 & 1.5299 & \textbf{0.0259} & 1.4013 & 18.5444  \\ \hline
 benzene & \textbf{2470} & 7602 & 3356 & 3267 & 4738  & \textbf{1481} & 7477 & 2321 & 3237 & 4092  & 18.4479 & 13.607 & \textbf{4.6286} & 15.9836 & 210.8994 \\ \hline

 LABS-(n10)  & \textbf{106} & 296 & 116 & 230 & 145  & \textbf{76} & 173 & 77 & 209 & 127  & 1.608 & 0.1553 & \textbf{0.0199} & 1.9484 & 4.2117  \\ \hline
 LABS-(n15) & \textbf{385} & 1208 & 457 & 880 & 641  & \textbf{255} & 672 & 281 & 797 & 568  & 21.797 & 0.5156 & \textbf{0.1077} & 0.4199 & 21.3482 \\ \hline
 LABS-(n20)  & \textbf{1052} & 2914 & 1138 & 2218 & 1762  & 679 & 1731 & \textbf{677} & 2023 & 1673  & 210.1475 & 1.4367 & \textbf{0.5801} & 1.2457 & 64.2246 \\ \hline
 MaxCut-(n15, r4) & 68 & \textbf{58} & 94 & 60 & 62  & 32 & 28 & 50 & \textbf{26} & 36  & 1.0402 & 0.0678 & 0.0399 & \textbf{0.0299} & 2.092 \\ \hline
 MaxCut-(n20, r4)  & 88 & \textbf{78} & 126 & 80 & 100  & 34 & 28 & 60 & \textbf{22} & 46  & 2.1981 & 0.0928 & 0.0848 & \textbf{0.0399} & 3.6862 \\ \hline
 MaxCut-(n20, r8) & \textbf{129} & 158 & 188 & 160 & 210  & 59 & 94 & 104 & \textbf{48} & 123  & 5.3028 & 0.1436 & \textbf{0.0888} & 0.1074 & 7.6895 \\ \hline
 MaxCut-(n20, r12) & \textbf{172} & 238 & 218 & 240 & 247  & 93 & 158 & 112 & \textbf{70} & 121  & 11.0385 & 0.2613 & \textbf{0.1207} & 0.1636 & 8.1801 \\ \hline
 MaxCut-(n10, e12) & 26 & \textbf{22} & 33 & 24 & 24  & 21 & \textbf{10} & 18 & 12 & 12  & 0.2882 & 0.0319 & \textbf{0.009} & 0.016 & 0.843 \\ \hline
 MaxCut-(n15, e63) & \textbf{93} & 114 & 108 & 102 & 137  & 51 & \textbf{36} & 59 & 42 & 69  & 2.6584 & 0.6096 & \textbf{0.0389} & 0.0904 & 4.4406 \\ \hline
 MaxCut-(n20, e117) & \textbf{146} & 216 & 188 & 192 & 298  & 65 & \textbf{56} & 85 & \textbf{56} & 121  & 8.4899 & 0.1715 & \textbf{0.0918} & 0.2055 & 10.1784 \\ \hline
\end{tabular}
    \begin{tablenotes}
    \normalsize
    \item The results for both QuCLEAR and PH were obtained after applying Qiskit optimization level 3. The compile time represents the total time required for running both QuCLEAR and the Qiskit optimization. PH denotes Paulihedral. 
    \end{tablenotes}
\end{threeparttable}
}
\end{table*}
\section{Evaluation}
\label{sec:evaluation}
\subsection{Comparison with the State-of-the-art Works}
Table~\ref{table:FC_results} shows the CNOT gate count, entangling depth, and the compilation time on a fully connected device. The best results are marked in bold. In summary, QuCLEAR can reduce the CNOT gate count by up to 68.1\% (50.6\% average), 52.5\% (31.5\% average), 63.9\% (43.1\% average), and 77.7\% (49.3\% average) compared with Qiskit, Rustiq, Paulihedral, and $\text{T}\vert \text{ket} \rangle$, respectively. QuCLEAR can reduce the entangling depth by up to 80.2\% (58.7\% average), 53.6\% (36.0\% average), 76.2\% (51.2\% average), and 84.1\% (59.1\% average) compared with Qiskit, Rustiq, Paulihedral, and $\text{T}\vert \text{ket} \rangle$, respectively. The average improvements are calculated as geometric means.

\textbf{CNOT Gate Count} QuCLEAR produces the fewest CNOT gates in 16 out of 19 benchmarks. The reduction in CNOT gates is particularly significant for the chemistry simulation benchmarks. This is because these benchmarks involve more complex Pauli strings, whereas the combinatorial optimization problems have Pauli strings in a more uniform format. Our recursive tree synthesis algorithm provides a greater advantage in optimizing the complex Pauli strings. The three cases where QuCLEAR does not produce the best results are the MaxCut problems with relatively sparsely connected graphs. In our optimization process, extracting the Cliffords through the mixer layer converts the single-qubit Rx rotations into multi-qubit rotations, as shown in Figure~\ref{fig:QAOA_opt}(b).  In other words, while we simplify the circuit in the problem Hamiltonian layer, we inadvertently increase the complexity in the mixer layer. Therefore, our approach is more advantageous for more complex problems, such as MaxCut graphs with higher-order connections or the LABS problem. For the 20-qubit MaxCut problems with 8- or 12-regular graphs, our compiler is the only one that achieves a non-trivial reduction in CNOT gate count compared with the original circuit.


\textbf{Entangling Depth} QuCLEAR produces the circuits with shortest entangling depth for all the chemistry simulation benchmarks and two of the LABS benchmarks, with the third LABS comparison being very close to the best. This result aligns with our previous discussion. Although minimizing circuit depth is not our primary target, extracting the Clifford subcircuit inherently reduces the circuit depth. For instance, in a standard V-shaped quantum simulation circuit, extracting half of the CNOTs effectively cuts the circuit depth in half. The extraction of the Clifford circuits and the consequent reduction in the CNOT gate count contribute significantly to the decreased entangling depth. For the MaxCut benchmarks, however, increasing the complexity in the mixer layer leads to a greater circuit depth and may result in long chains of CNOT gates, which can increase the overall circuit depth.

\textbf{Compile Time} Among all the methods, Rustiq has the shortest compile time for most of the benchmarks, owing to its implementation in Rust~\cite{klabnik2023rust}. In contrast, our implementation is in Python. For the chemistry simulation benchmarks, the runtime of QuCLEAR with Qiskit optimization level 3 is very close to the runtime of Qiskit alone. The significant reduction in circuit size shortens the runtime for Qiskit's optimization. 
Notably, for the largest benchmark, UCC-(10,20), QuCLEAR achieves the best compile time, being orders of magnitude faster than Rustiq. For the QAOA problems, QuCLEAR experiences longer compile times due to the search required within the commuting blocks. Nevertheless, it still manages to achieve compile times comparable to $\text{T}\vert \text{ket} \rangle$.

\begin{figure}[htbp]
\centerline{\includegraphics[width=\linewidth]{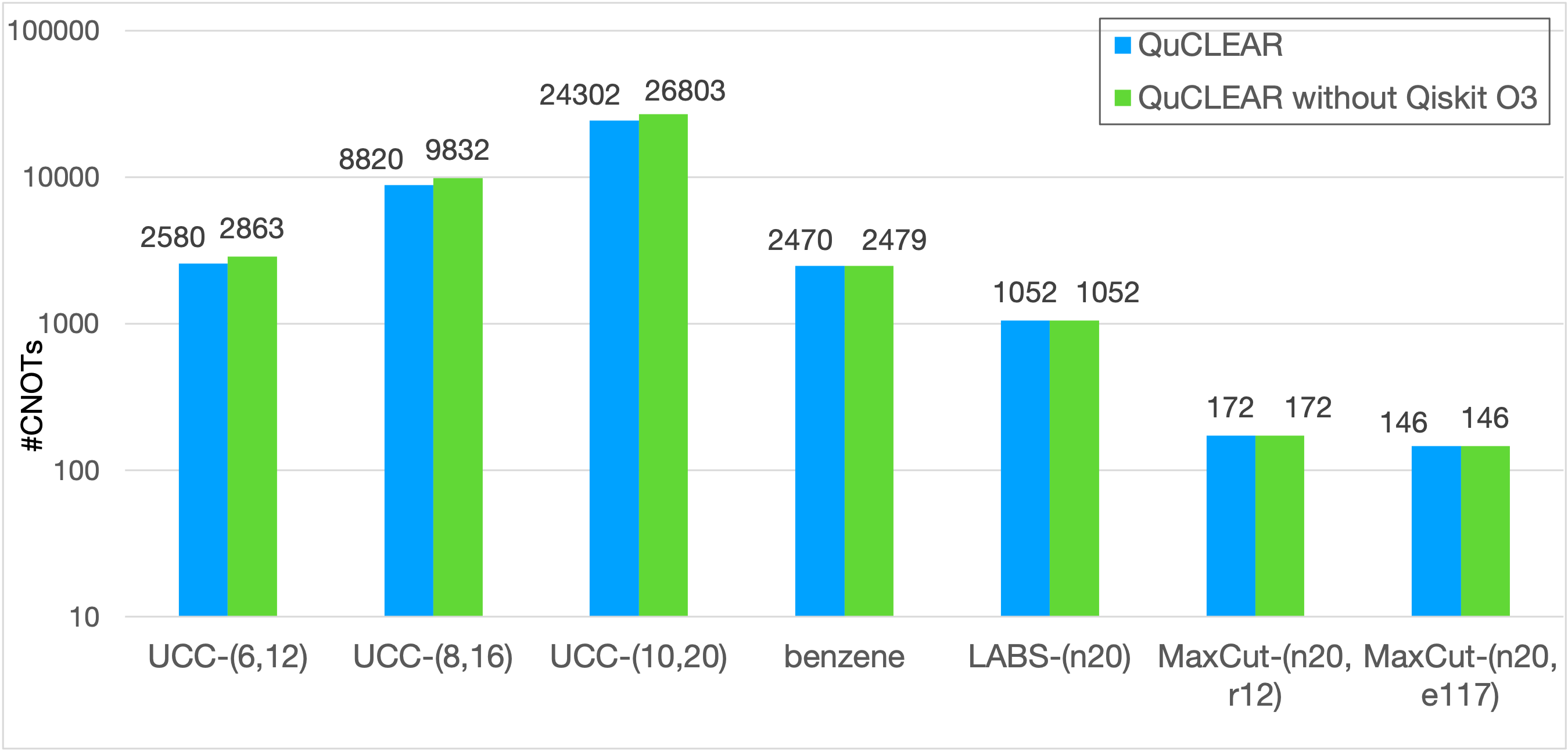}}
\caption{Comparison of QuCLEAR with and without Qiskit optimization}
\label{fig:without_opt} 
\end{figure}

We also present the QuCLEAR results and runtime without applying Qiskit optimization. As shown in Figure~\ref{fig:without_opt}, the Qiskit optimization does not reduce the CNOT counts for the QAOA benchmarks. Applying Qiskit optimizations results in an average CNOT count reduction of $4.4\%$ and increases the compile time by $6\%$. This confirms that our framework is effective on its own.

\subsection{Clifford Absorption Runtime}
\begin{table}[htbp]
\centering
\caption{Clifford Absorption runtime (s) for UCC-(10,20) and MaxCut-(n20, r12) benchmarks}
\label{table:CA_runtime}
\resizebox{\linewidth}{!}{%
\begin{tabular}{|c|c|c|c|c|c|c|}
\hline
Number & 10 & 50 & 100 & 500 & 1000 & 5000\\ \hline
Observables & 0.047 & 0.108 & 0.210 & 1.071 & 2.189 & 12.161\\ \hline
States  & 0.002 & 0.009 & 0.015 & 0.079 & 0.156 & 0.585\\ \hline
  
\end{tabular}
}
\end{table}
The runtime of the Clifford Absorption (CA) module scales linearly with the number of observables or states. We measured the runtime separately because it depends not only on the circuit but also on the number of observables or states considered. To illustrate this, we selected the largest chemistry simulation benchmark, UCC-(10, 20), and the numerical optimization benchmark, MaxCut-(n20, r12). 
We executed the CA module with varying numbers of observables and states. As shown in Table~\ref{table:CA_runtime}, the runtime scales linearly. For a reasonable number of observables and states, the runtime remains acceptable compared with the overall compile time.

\subsection{Closer Examination of the Improvements}
\begin{figure}[htbp]
\centerline{\includegraphics[width=\linewidth]{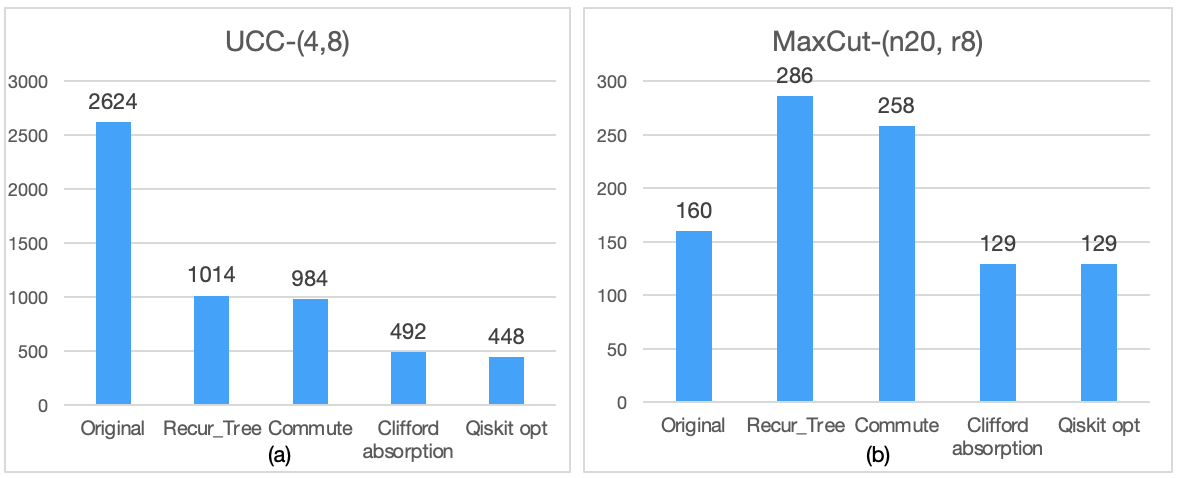}}
\caption{CNOT gate reduction achieved by each individual feature in optimizing the UCC-(4,8) and MaxCut-(n20, r8) benchmark.}
\label{fig:UCC_breakdown} 
\end{figure}

Since we have introduced several optimizations, we must quantify the contributions. We selected two benchmarks, UCC-(4,8) from the chemistry simulation category and MaxCut-(n20, r8) from the combinatorial optimization category. In Figure~\ref{fig:UCC_breakdown} we present a detailed breakdown of the CNOT gate count reduction achieved by each optimization. 

For the chemistry simulation problems, the Pauli strings contain many non-identity Pauli operators. As a result, recursively synthesizing the CNOT tree and extracting the Clifford subcircuit significantly optimize the circuit, reducing the gate count from 2,624 to 1,014. Since few Pauli strings commute with their neighbors, incorporating commuting considerations only slightly reduces the gate count to 984. The next notable reduction occurs when we absorb the Clifford subcircuits, effectively halving the gate count. And with the local rewriting optimizations in Qiskit, the final gate count is reduced to 448.

The QAOA for MaxCut benchmark exhibits a different pattern. It consists of Pauli strings with only one or two non-identity Pauli operators. Therefore, simply extracting the Clifford circuits can potentially increase the circuit size, since it often results in a greater number of non-identity Pauli operators in the strings. Since many Paulis commute, the commutation optimization reduces the CNOT gate count from 286 to 258. Absorbing the CNOT network also halves the gate count, bringing it down to 129. The subsequent optimizations in Qiskit do not further reduce the gate count.

\subsection{Mapping to Devices with Limited Connectivity}
\begin{figure}[ht]
    \centering
    \subfloat[Mapping to Google Sycamore]{
        \centering
        \includegraphics[width=0.9\linewidth]{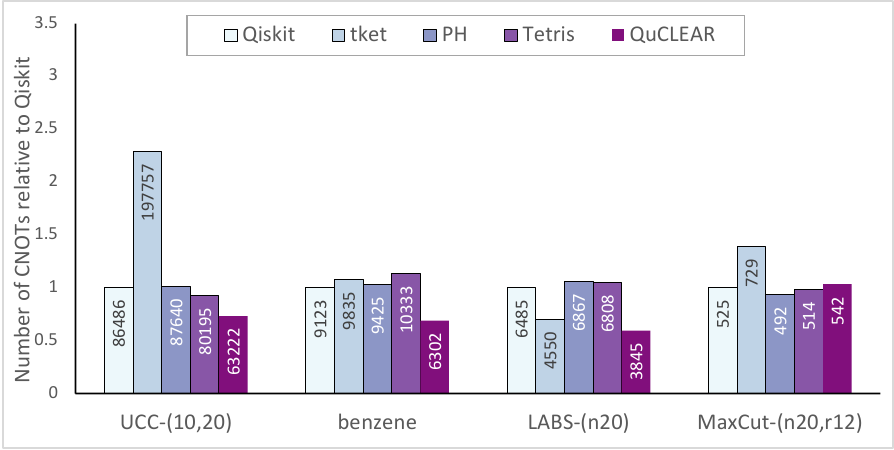} 
        \label{fig:subfig1}
    }\\
    \subfloat[Mapping to IBM\_Manhattan]{
        \centering
        \includegraphics[width=0.9\linewidth]{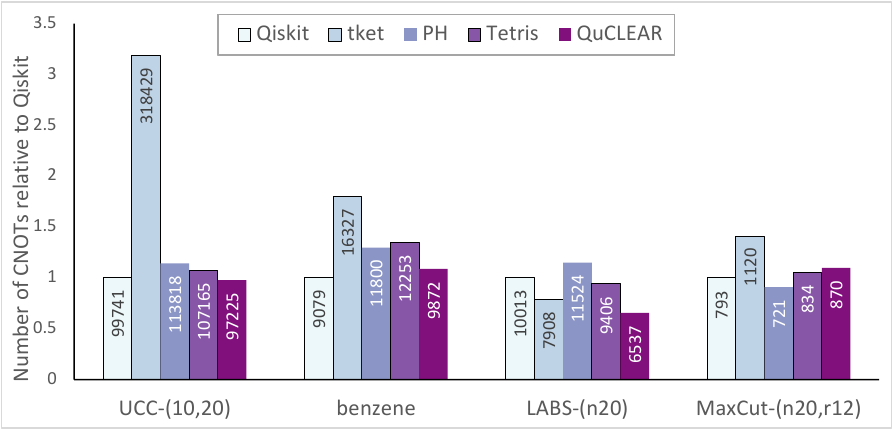} 
        \label{fig:subfig2}
    }
    
    \caption{Comparison of mapping to different backends}
    \label{fig:sparse_backend}
\end{figure}

We evaluated QuCLEAR's performance by mapping the optimized circuit to two quantum devices with limited connectivity: Google Sycamore and IBM\_Manhattan. Figure~\ref{fig:sparse_backend} presents our experimental results using Qiskit’s highest optimization level for device mapping. Due to space constraints, we report the largest benchmark from each circuit type. QuCLEAR produced the smallest circuit in 3 out of 4 benchmarks on Sycamore, with average CNOT gate reductions of $25.4\%$, $40.1\%$, $26.1\%$, and $26.8\%$  compared to Qiskit, Tket, Paulihedral, and Tetris, respectively. Since Rustiq’s results do not include single-qubit rotation gates, we have excluded it from the mapping comparison. 
When mapping to IBM\_Manhattan, QuCLEAR results in the smallest circuit in 2 out of 4 benchmarks, with average reductions of $6.7\%$, $41.3\%$, $16.4.1\%$, and $14.7\%$. These results demonstrate that QuCLEAR remains effective even on devices with sparse connectivity. However, the heavier insertion of SWAP gates when mapping to IBM\_Manhattan’s more limited heavy-hex topology led to reduced gate count improvement.
It’s important to note that while Paulihedral and Tetris are hardware-aware, QuCLEAR is designed at the logical circuit level. With improved mapping techniques, further reductions in gate count could be achieved.

\section{Conclusion}
In this paper we propose QuCLEAR, a compilation framework for optimizing quantum circuits. We present two distinct techniques: Clifford Extraction and Clifford Absorption. We propose an efficient heuristic algorithm for synthesizing CNOT trees in quantum simulation circuits. Our approach leverages classical devices to efficiently simulate portions of the computations, leading to a reduced circuit size on quantum devices. Experimental results across various benchmarks demonstrate that QuCLEAR outperforms state-of-the-art methods in terms of both CNOT gate count and circuit depth.
\label{sec:conclusion}

\section*{Acknowledgments}
We thank the anonymous reviewers for their detailed feedback and express our gratitude to Gail Pieper for her assistance in revising the paper. This material is based upon work supported by the U.S. Department of Energy, Office of Science, National Quantum Information Science Research Centers. This material is also based upon work supported by the DOE-SC Office of Advanced Scientific Computing Research AIDE-QC project under contract number DE-AC02-06CH11357.



\bibliographystyle{IEEEtranS}
\bibliography{refs}

\newpage

\appendix

\section{Artifact Appendix}

\subsection{Abstract}

Our artifact includes the source code for QuCLEAR and Python scripts for generating optimized circuits for the benchmarks. These scripts also produce output necessary for validating the results. Users can reproduce the key findings presented in Tables~\ref{table:FC_results},~\ref{table:CA_runtime}, and Figures~\ref{fig:without_opt},~\ref{fig:UCC_breakdown},~\ref{fig:sparse_backend} using the provided scripts.

\subsection{Artifact check-list (meta-information)}

{\small
\begin{itemize}
  \item {\bf Algorithm: } QuCLEAR's recursive tree synthesis algorithm is discussed in Algorithm~\ref{alg:Tree_synthesis}.
  \item {\bf Program: } Qiskit, TKet, Rustiq, Paulihedral(PH), Tetris
  \item {\bf Compilation: } Python
  \item {\bf Data set: }Benchmarks listed in Section~\ref{sec:methodology}. Series of Pauli strings from self-generated UCCSD, Hamiltonian simulation, LABS and MaxCut dataset.
  \item {\bf Run-time environment: } macOS, Linux, or Windows, and Python
  \item {\bf Hardware: } Server or PC.
  \item {\bf Execution: } Run the bash scripts and python scripts
  \item {\bf Metrics: }CNOT gate count, circuit depth, compile time
  \item {\bf Output: }JSON files and tables in jupyter notebook
  \item {\bf Experiments: } Compare the metrics on the provided datasets
  \item {\bf How much disk space required (approximately)?: }5GB
  \item {\bf How much time is needed to prepare workflow (approximately)?: }1 hour
  \item {\bf How much time is needed to complete experiments (approximately)?: }12 hours
  \item {\bf Publicly available?: }Yes
  \item {\bf Code licenses (if publicly available)?: }Apache-2.0 License
  \item {\bf Archived (provide DOI)?: } 10.5281/zenodo.14353198
\end{itemize}
}

\subsection{Description}

\subsubsection{How to access}

The source code for QuCLEAR and the scripts to run the benchmarks are available in github: \url{https://github.com/revilooliver/QuCLEAR}. We recommend download from github to get the most recent updates.
\subsubsection{Hardware dependencies}
You need a regular PC with an Intel or AMD CPU.
\subsubsection{Software dependencies}
Python 3.10, Qiskit 1.1.1, pytket, Paulihedral, and rustiq are used in our experiments.  
\subsubsection{Data sets}
Benchmarks are listed in Section~\ref{sec:methodology}. We provide the self-generated dataset in `benchmarks' folder so re-generating the benchmarks is not necessary. However, if you wish to generate them yourself, the necessary scripts are also included in the same folder.

\subsection{Installation}

\subsubsection{Install conda}
Download Anaconda at https://www.anaconda.com/ and install it.

\subsubsection{Download the Repository}
First, clone the QuCLEAR repository (or download from the zenodo repo) and navigate to its directory:

\$ git clone https://github.com/revilooliver/QuCLEAR.git

\$ cd QuCLEAR



\subsubsection{Virtual Environment Setup}
For artifact evaluation, QuCLEAR is compared with several tools: Rustiq, Qiskit, Paulihedral, pytket and Tetris. To simplify installation and environment management, we provide an automated script, \texttt{install\_all.sh}, located in the `artifact\_evaluation' folder:

\$ cd artifact\_evaluation

\$ ./install\_all.sh

This script creates three Anaconda virtual environments: QuCLEAR\_env, PH\_env, and Tetris\_env. QuCLEAR\_env contains QuCLEAR, Qiskit, and pytket. PH\_env contains Paulihedral. Tetris\_env contains Tetris.

\subsubsection{Install Rust and Rustiq} After setting up the environments, you will need to manually install Rust and Rustiq in the QuCLEAR\_env.

First, activate the QuCLEAR environment:

\$ conda activate QuCLEAR\_env

Then follow the installation instructions provided on the official Rust website to install Rust: \url{https://www.rust-lang.org/tools/install}

After installation, source your .bashrc file to enable Rust in your current session:

\$ source ~/.bashrc

You may need to reactivate the environment to ensure the paths are properly configured:

\$ conda deactivate

\$ conda activate QuCLEAR\_env

Finally, clone and install Rustiq from its GitHub repository:
\url{https://github.com/smartiel/rustiq/tree/main}

Once these steps are complete, all required dependencies for artifact evaluation will be ready.

\subsubsection{Test Installation}
We provide a simple test script to validate the installation. It will compile the circuits for Hamiltonian simulation:

\$ ./test\_experiments\_fast.sh

\subsection{Experiment workflow}
\subsubsection{Generate compilation results}
The experimental data can be generated by running the full experiment script:

\$ cd artifact\_evaluation

\$ ./test\_experiments\_full.sh

The compilation results are saved as JSON files in the `experiments/results\_fullyconnected' folder.

For example, `test\_quclear\_LiH.json' contains results for QuCLEAR, Qiskit, and Rustiq for the LiH benchmark. In the JSON file, \textbf{our\_method} indicates results using QuCLEAR and \textbf{combined\_method} indicates results using QuCLEAR combined with Qiskit's optimization (reported in the paper).

The hardware mapping experimental data in Figure~\ref{fig:sparse_backend} can be generated by running the following experiment script:

\$ cd artifact\_evaluation

\$ ./test\_mapping\_experiments\_full.sh

The compilation results are saved as JSON files in the `experiments/results\_ibm' and `results\_google' folders.

\subsubsection{Generating Table Data}
To generate the experimental tables in our paper, use the provided Jupyter notebooks generate\_table3.ipynb, and generate\_table4.ipynb in the `artifact\_evaluation' folder. These notebooks process results generated from the previous step that are stored in the `experiments/results\_fullyconnected' folder.

To generate the figuers in our paper, use the provided Jupyter notebooks generate\_figure9, 10 and 11.ipynb in the `artifact\_evaluation' folder. The figures are stored in the `/artifact\_evaluation/figures' folder.

\subsection{Evaluation and expected results}
The gate counts and circuit depth in the jupyter notebooks should closely match the results in the Tables and Figures. However, due to the inherent randomness in Qiskit's compiler optimization, the gate counts and depths may fluctuate slightly. While compile times may vary, the overall trends should remain consistent.

\subsection{Experiment customization}

The benchmark\_comparison\_QuCLEAR\_Qiskit\_rustiq.py file in `experiments' folder can be modified to run different benchmarks and change the configuration of the algorithm.

Our two examples in tutorial folder provides example of running the whole evaluation of the circuit and measure the results with a simulator. 

\subsection{Methodology}

Submission, reviewing and badging methodology:

\begin{itemize}
  \item \url{https://www.acm.org/publications/policies/artifact-review-and-badging-current}
  \item \url{http://cTuning.org/ae/submission-20201122.html}
  \item \url{http://cTuning.org/ae/reviewing-20201122.html}
\end{itemize}

\end{document}